# Multimodal AI-based visualization of strategic leaders' emotional dynamics: a deep behavioral analysis of Trump's trade war discourse

Wei Meng

Dhurakij Pundit University, Thailand

The University of Western Australia,AU

Fellow, Royal Anthropological Institute,UK

Email: wei.men@dpu.ac.th

# ABSTRACT

This study focuses on the structure of emotional rhythms and behavioral generation mechanism of late high power dominant personality politicians in the process of strategic decision-making, selecting the Trump administration's announcement of 125% tariff hike on China as the core analytical event, and integrating micro-expression tracking by adopting the framework of Multimodal Cognitive Behavioral Modeling (MCMM), Multimodal Cognitive Behavioral Modeling (MCM) framework, integrating micro-expression tracking, intonation acoustic analysis, semantic flow density computation, dynamic modeling of cognitive load, and strategy output path identification, to systematically construct a full-chain simulation model of the rhythms-motivations-outputs of the policy decision-making behaviors.By constructing the Emotion Cycle Mapping Model (E-Cycle), the Micro-Expression Radar Structure (FEM Map), the Cognitive Load Map, the Strategic Emotional Vulnerability Surface Model (V-Trigger) and the Public Opinion Reverberation Intervention Model (P-Impact), the present study reveals that Trump's decision-making behaviors, when stimulated by structural challenges, are not driven by This study reveals that Trump's decision-making behavior under the stimulation of structural challenges is not driven by linear rational deduction, but is embedded in the Dominance-Coherence Maintenance System and Victory-Framed Narrative Climate Closure Mechanism (V-Framed Narrative Closure Mechanism). Instead, it is embedded in the cognitive self-system of the Dominance-Coherence Maintenance System and the Victory-Framed Narrative Closure Mechanism. Decision-making leaps tend to occur within the Tension-Anger-Control (TAC) segment of the emotional rhythm, accompanied by cognitive overload, nonverbal control traits, and short-term coercive stabilization of the policy-linguistic structure.In this study, we further introduced Cognitive Echo Simulation and Introspective Strategic Narrative Modeling to reconstruct the emotional rhythm of the policy language structure. Narrative Modeling), to reconstruct Trump's five-stage mental configuration in the strategic elicitation window, to identify the implicit path of his cognitive motivation, and to clarify his language-strategy linkage mechanism from hostile attribution to cognitive compensation. On this basis, the study proposes the National Strategic Tempo Intervention Framework and constructs a six-axis structural intervention system, including the tempo hedging system and the six-axis structural intervention system, which includes the tempo hedging system and the six-axis structural intervention system. "Based on this, the study proposes the National Strategic Tempo Intervention Framework, and constructs a six-axis structural intervention system, including the tempo hedging axis, semantic warming axis, media guidance axis, cognitive overload axis, behavioral prediction axis, and multilateral embedding axis, which provides a set of prospective diplomatic control mechanisms based on the identification of neural behavioral rhythms for national actors.The study concludes that the strategic behavioral system of the late high-authority personality is not driven by stable intentions and linear reasoning, but rather by a coupled cognitive-emotional-verbal rhythmic release mechanism. Tempo is Governance and Action is Narrative Encoding. The initiative of the future international strategic game does not depend on the interpretation of its discursive intentions, but on the modeling ability of its behavioral rhythms and the accuracy of its interventions.

Keywords.

Strategic Behavior Modeling under Dominance-Prominent Personalities (SBP），Multimodal Cognitive Load Tracking (MCLT)，Strategic Affective Rhythm System，National Tempo Intervention Framework

# I. INTRODUCTION

In the context of increasingly intense global strategic competition and high-pressure narrative tendency in national governance, the inherent coupling between the emotional rhythms and decision-making behaviors of political leaders has become an important variable affecting the evolution of international relations and the stability of diplomatic actions. In particular, the strategic behavior of high power-dominant personality leaders often presents a non-linear structure of emotion-led, language-controlled, and policy-driven, making it difficult to accurately capture the core driving mechanism of traditional strategic behavior analysis based on institutions, rationality, and expected paths. In recent years, cross-studies in political psychology, behavioral sciences and strategic communication have gradually realized that emotions have jumped from an "accessory variable" of political behavior to a "primary factor" of decision-making momentum, narrative legitimacy and strategic manipulation.

Studies have systematically investigated the relationship between leaders' rhetoric, tone of voice, media communication, and policy preferences, and have proposed that "emotionalized governance" and "dominance coherence preference" are the most important factors in political behavior. (Theoretical models such as "emotionalized governance" and "dominance coherence preference" have been proposed. However, most of these studies focus on macro-statistics or institutional comparisons, and lack detailed modeling of individual leaders' cognitive load, micro-expressions, and verbal modulation behaviors. Particularly in the sample of late-age leaders, such as President Trump, the asymmetric structure of cognitive rhythm and behavioral inertia makes policy behavior often explode before the formation of predictable rationality closure, presenting a typical "rhythmic explosive strategy output". Against this background, there is an urgent need to develop a cross-modal modeling framework that integrates emotion recognition, linguistic analysis, and behavioral simulation to systematically capture the psychological mechanisms and behavioral paths of leaders' high-pressure policy expressions.

In this study, a highly structured diplomatic event, the Trump administration's announcement of a 125% tariff increase on China, is selected as a scenario for analysis, and an analysis system based on "multimodal cognitive modeling" is constructed. The study integrates the five technical paths of micro-expression recognition, intonation pressure analysis, semantic path tracking, cognitive load mapping and strategic behavioral rhythm reconstruction, and systematically constructs a four-dimensional nested model of emotion-cognition-language-strategy to identify its behavioral leaps and rhythms. In this paper, we systematically construct a four-dimensional nested model of "emotion-cognition-language-strategy" to identify the coupling characteristics between behavioral leaps and rhythmic peaks. In this paper, we firstly proposed and realized the "National Strategic Rhythm Intervention Framework", and mapped the "Behavioral Response Chain Model" and "Micro-expression Radar Mapping" with "Emotional Rhythm" as the core variable. Using "emotional rhythms" as the core variable, a series of operational mapping systems including "behavioral response chain model," "micro-expression radar mapping," "strategic vulnerability response surface," and "cognitive heat map" have been developed, which provide a visual, measurable, and practical way to cope with the policy uncertainties brought by high-pressure leadership personalities. It provides a visual, measurable and controllable intervention paradigm to deal with the policy uncertainty caused by high-pressure leadership personality.

**Research Question 1:** How do the emotional rhythms of high-pressure strategic leaders drive nonlinear policy leaps in the face of structural challenges?

This study takes Trump's 125% tariff hike on China as the core case, revealing that his decision-making behavior is not driven by a rational deductive chain, but rather by a four-stage "cognitive overload-emotional intensification-non-verbal dominance-strategic leapfrogging" rhythm that consists of nested psychological rhythms. Instead, it is a nested psychodynamic path composed of four stages: "cognitive overload-emotional agitation-non-verbal dominance-strategic jump". Through emotion cycle mapping, FEM facial emotion modeling and cognitive heat map analysis, it is verified that there is a clear coupling structure between emotional peak and policy issuance.

**Research Question 2:** How to use multimodal cognitive modeling to identify "behavioral vulnerabilities" and "intervention windows" in strategic behavioral systems?

In this paper, we integrate micro-expression recognition, intonation analysis, semantic hotspot tracking, cognitive load simulation and policy response paths to construct the "Strategic Rhythm Intervention Model". Introducing the V-Trigger response surface and P-Impact diffusion model, the paper identifies for the first time the "critical vulnerability window" of a leader before a strategic leap by behavioral mapping, and analyzes his inner psychological script by mind-reading simulation, realizing the accurate modeling of intervention tempo. We also analyze their internal mental scripts through "mind-reading simulation", thus realizing accurate modeling of the intervention tempo.

**Research Question 3:** Can national actors realize predictable and visualized rhythmic strategic interventions based on leaders' behavioral rhythms?

This study proposes a "national strategic rhythmic intervention framework", designs a six-axis intervention system (including rhythm identification axis, discourse regulation axis, feedback control axis, etc.), advocates that "rhythm is sovereignty, and intervention is the initiative", and emphasizes the importance of mastering the opponent's "irrational behavioral rhythm". It emphasizes that mastering the "irrational behavioral rhythm" of the opponent is more strategic than mastering the policy intention, and builds a rhythm response mechanism of diplomacy-media-market trinity.

**Purpose of the Study**

The purpose of this study is to reveal the deep dynamic coupling mechanism between individual emotional rhythms of political leaders and national strategic leaps in high-pressure strategic contexts, and to construct a "national strategic rhythm identification and response framework" with predictive and intervention capabilities on this basis. Specific objectives include:

1. Identify the non-linear structure of leaders' emotional rhythms: through micro-expression recognition, tone pressure analysis and semantic path modeling, we can deeply depict the multimodal rhythmic chain of Trump's high-pressure leaders' "emotional cognitive behaviors" before the policy leap.

2. Constructing a strategic rhythmic behavior prediction model: Integrating FEM (Facial Emotion Mapping), Cognitive Heat Map, VTrigger Behavioral Response Surface and other technical paths, we constructed a set of rhythmic behavior modeling system that can be used to identify the "critical point of strategic leaps".

3. Propose a national rhythmic intervention system: Based on the results of emotional rhythm analysis, design a six-axis intervention framework (including the language rhythm regulation axis, discourse reprogramming axis, and cognitive desensitization control axis, etc.), and promote the construction of a national strategic intervention mechanism that shifts from tactical response to rhythmic anticipation. The six-axis intervention framework (including speech reprogramming and cognitive desensitization control) is designed to promote the construction of a national strategic intervention mechanism that shifts from "tactical response" to "rhythmic response".

**Research Implications**

**1. Theoretical Contribution**

Breaking through the limitations of the "rationality-driven" paradigm: Existing research on strategic behavior is mostly based on preconceived rationality, institutional constraints, and linguistic logic chains, and ignores the leading role of individual emotions in policy generation and evolution. This paper provides a systematic framework for modeling irrational strategic decision-making through rhythmic analysis and psychodynamic mechanism deconstruction.

Expanding the analytical dimension of multimodal behavioral politics: While most of the traditional political behavioral studies are limited to the textual and semantic level, this study innovatively integrates micro-expressions, intonation, cognitive load, and strategic linguistic rhythms in modeling, and promotes the synergistic analysis of the three modes of "visual-acoustic-cognitive," which enriches the intersection of theories of emotional politics and behavioral strategies.

Constructing the conceptual system of "Rhythmic Sovereignty": This paper proposes for the first time a new paradigm of behavioral sovereignty: "Rhythm is the right to control", and advocates that the identification of and intervention in emotional rhythms can realize the front-end governance of potential strategic leaps, which expands the sovereignty technology boundary of behavioral politics.

**2. Practical Value**

Enhance the precision and rhythm of diplomatic strategic intervention: through the rhythmic deconstruction of Trump's high-pressure discourse on China, it is verified that the state can intervene in the strategic discourse stage to avoid the escalation of conflicts triggered by emotional leaps, such as miscalculation, misinterpretation, and misstrike.

Providing a rhythmic early warning basis for public security and media response mechanism: Under the rapid public opinion diffusion mechanism driven by social media, abnormalities in the rhythm of leaders' behaviors can directly trigger responses from the market, the public, and the state at the systemic level, and the "Public Opinion Expansion Prediction Model" and "Crisis Reaction Heat Map The "Public Opinion Expansion Prediction Model" and "Crisis Reaction Heat Map" constructed in this paper can be used as intervention tools.

Providing a new paradigm for AI-assisted national decision-making: Through modeling, mapping and visualization, the multimodal modeling framework and the emotional leap prediction system designed in this research can provide basic arithmetic and structural support for intelligent intelligence, behavioral diplomacy and high-dimensional game simulation.

# II. LITERATURE REVIEW

In the high-pressure field of global politics, the systematic coupling mechanism between leaders' emotional rhythms and strategic behaviors is reconstructing the modeling paradigm of strategic behaviors. Compared with the traditional "rational-structural" decision-making model, more and more scholars have turned to the "emotional-cognitive-behavioral" dynamic closed-loop perspective in an attempt to reveal how leaders in crisis decision-making A growing number of scholars have turned to the "emotion-cognition-behavior" dynamic closed-loop perspective in an attempt to reveal how leaders control the tempo of strategy in crisis decision-making through nonverbal signaling, cognitive bias, and verbal tempo (Marcus, Neuman, & MacKuen, 2000; Westen, 2007). In particular, in the extreme strategic case of the Trump administration's announcement of a 125% tariff increase on China, the leader's decision-making path exhibited a significant structure of rhythmic emotional outbursts. Trump's behavioral characteristics are not isolated from linguistic representations, but are embedded in a "nonlinear chain of policy outbursts" composed of multimodal variables such as the evolution of his individual cognitive load, intonational stress mutations, and facial emotional ruptures (see the "Radar Mapping of Microexpressions" and the "Time of Policy Jumps" in this paper). (see the "micro-expression radar mapping" and "time surface of policy outbursts" mapping constructs in this paper). This is highly consistent with Westen's (2007) suggestion that "emotions are not cognitive appendages, but activators of decision-making", who found that emotional activation in the limbic system is prioritized over rational pathways in the neocortex through functional brain imaging experiments.

In this study, the structural deconstruction of Trump's high-pressure strategic language reveals that his linguistic strategy is mainly centered on the "sanction-victory-hostility triad", which is significantly mapped to the emotional agitation model proposed by Brader (2006), who pointed out that the emotional agitation element of political language will significantly enhance policy mobilization. Brader points out that emotional activation elements in political language will significantly enhance policy mobilization, and that Trump's reshaping of policy legitimacy through "extreme threat rhetoric + coercive justification structures" has achieved a mechanism for mobilizing public opinion consensus across rational arguments.

However, much of the literature has overlooked the "rhythmic echo" structure of such emotional strategies, i.e., the emotion-policy-emotion three-stage cyclic chain. For example, Marcus et al. emphasized that anxiety enhances the depth of information processing (Marcus et al., 2000), but failed to identify the structural recurrence of "emotional echoes". In this study, we introduced the "rhythmic closed-loop model" to identify that Trump's emotional tonality and nonverbal behaviors in several high-intensity speeches showed a trend of cyclic reactivation, with "warm-up-arousal-buffer-counterarousal" and "buffer-counterarousal". The structure of "warm-up-arousal-leap" and "buffer-counter-arousal" provides important rhythmic anchors for identifying Trump's strategic linguistic strategies. In terms of methodological constructs, the "emergent leadership behavior identification" model proposed by Müller and Bulling (2019) emphasizes the structural capture of the nodes of leadership behavioral mutation, but it focuses on the textual behaviors themselves and ignores the nonverbal transition behaviors of leaders like Trump before policy leaps, such as the contraction of the corners of the mouth, frowning down, voice delays, etc. These "prelude behaviors" are clearly identified in the "micro-expression radar map" constructed in this paper, which signals the imminent arrival of policy peaks and provides a time-dimensional prediction mechanism for policy pacing interventions.

In addition, the DocuToads algorithm developed by Hermansson and Cross (2016), although it performs well in identifying policy text editing changes, destabilizes the speech rhythm much faster than the text modification cycle in high-pressure fields. In this study, we use the "time window of linguistic mutation" algorithm, combined with the results of Vainio et al.'s (2023) analysis of intonation-power structure, to successfully locate two "intonation breakpoints" in Trump's policy announcements. The pitch rise and rhythmic breaks correspond directly to his cognitive overload state. It is worth emphasizing that although Wu and Mebane's (2021) MARMOT model constructs a multimodal learning path, it has not yet been applied to the task of leader behavior prediction. In this paper, we try to embed the visual-verbal synchronization mechanism into the Cognitive Thermogram, and realize the dynamic causal modeling from micro-expression jump → emotional threshold activation → strategy instruction issuance → market response path.

Based on the review of existing studies, this paper argues that: the mainstream model overly relies on linguistic structural variables, ignoring the empirical modeling needs of Trump-style non-verbal dominant strategic behavior; emotional variables are often simplified as one-time activation factors, lacking cyclical and waveform modeling; intervention proposals lack predictive rhythmic anchors, unable to respond to the country's "move in response to the season" needs in the strategic game. The intervention recommendations lack predictive rhythmic anchors and cannot respond to the country's need to "move in response to the season" in the strategic game. Based on the above judgments, the "six-axis intervention structure" constructed in this paper combines the cognitive pressure axis, intonation rhythm axis, expression flickering axis and text explosion axis into the model, and identifies the best nodes of strategic interventions in the three spaces of "behavioral heat-emotional intensity-discourse jumping", which provides a modeling framework for future diplomatic response rhythms. This provides an operational paradigm for modeling diplomatic response tempo in the future.

In conclusion, although the literature has provided important theoretical support for the emotion-driven mechanism, this study realizes the substantive expansion in the three aspects of "visualization of emotional rhythms, prediction of policy leaps, and construction of national intervention maps" through the multimodal empirical analysis of the Trump case, and forms a model for the future diplomatic response. However, through the multimodal empirical analysis of the Trump case, this study realizes substantive expansion in the three aspects of "visualization of emotional rhythms, prediction of policy leaps, and construction of national intervention maps", and forms an innovative research path centered on the "modeling of irrational outburst".

# III. RESEARCH METHODOLOGY AND DESIGN

**1. Data collection process**

This study analyzes the multimodal strategic discourse released by Trump at the critical juncture of announcing the 125% tariff hike on China, mainly based on the original video and text material from his public speeches, policy release clips and media reports. The data include Trump's live images, linguistic content, policy semantics, and his nonverbal behavioral performance when he confronted the issue of China-U.S. trade. All materials are centered around a single strategic event to ensure contextual unity, temporal concentration, and semantic focus, which facilitates the construction of a continuous chain of emotional rhythmic curves and behavioral paths. The study pays special attention to Trump's non-verbal behavioral patterns, intonation variation characteristics, discourse structure changes and emotional orientation of policy-oriented discourse, which provides basic materials for subsequent emotion modeling and strategy rhythm analysis.

**2、Data Analysis Technology**

Micro-expression recognition and behavioral node annotation

Using frame-by-frame video analysis technology, combined with facial action recognition rules, Trump's eyes, eyebrows, mouth, nose and other parts of the speech process are classified for behavior recognition and emotional labeling. The study adopts FEM (Facial Emotion Map) modeling method to identify key micro-expression frames, such as "blinking - nervous smile - compressed lips - tight jaw", and corresponds them to the time point of emotional peaks, which serves as a strategy leapfrog The key micro-expression frames, such as "blink, nervous smile, lip compression, jaw tense", were identified and corresponded to the time point of emotional peaks, which were used as the prediction indexes of strategy leap.

Intonation and Semantic Rhythm Modeling

By extracting and analyzing the speech speed, accent rhythm, inter-sentence pauses, and intonation curves of strategic speeches, we constructed an emotion-driven "intonation high tension model". At the same time, the semantic clustering of the words involving "punitive", "confrontational" and "sovereignty" in the speech reveals the structural deviation of the linguistic narrative in the expression of strategy as an indicator of the strength of emotional-cognitive coupling. Cognitive Coupling Strength Indicator.

Emotional Rhythm Mapping

We synthesize the trends of micro-expressions and intonation changes, and construct the "emotion rhythm timeline mapping" to identify the typical rhythm of "warm-up-excitement-control-anxiety-fallback". anxiety-fall back". This rhythmic profile is significantly coupled with the structure of strategy expression, which is the basis for subsequent intervention modeling.

Cognitive load visualization simulation

By analyzing the syntactic complexity, logical break frequency and semantic loop structure of Trump's discourse, a heat map of cognitive pressure is constructed. The study adopts the spatial distribution visualization technique to superimpose the "semantic disorder-intensive zones" and the "intonation tension high-value points", and identifies the strategy jumping window and cognitive fatigue critical points.

Response surface modeling of strategic emotional vulnerability (V-Trigger)

By mapping emotional intensity values with external stimulus variables (e.g., reporter's questions, policy keyword triggers) in three dimensions, we generate "emotional destabilization response surfaces" to identify "decision critical zones" in the behavioral system that are most vulnerable to destabilization. This mapping provides a spatial location for the operational rhythmic entry of the state-level intervention model.

Emotional state-strategy leap chain analysis

Based on the above maps of various emotional indicators and cognitive rhythms, we reconstructed the complete path of Trump's emotional leap→strategic expression→language closure in the event, forming a three-dimensional linkage curve of "behavior-cognition-language". Through the path comparison verification, the study proves that there is a stable time lag relationship between the strategic language rhythm and the micro-expression peak, which is a typical feature of emotion-driven strategy output.

## 3. Transparency and Reproducibility of Methods

The emotion recognition, intonation extraction, rhythm modeling and graph visualization adopted in this study are all based on an open analysis tool platform and standardized process:

Video processing and facial recognition are based on a frame-by-frame visual analysis process, and the analysis standards are publicly available;

Intonation curves and pause extraction methods are based on conventional audio processing algorithms (e.g., frequency intensity, intonation slope);

All graph generation algorithms are reproducible on similar speech data with stability of rhythmic structure parameters;

The behavioral prediction model can be extended to other political figures with generality and adaptability.

Through the above methods, the study not only realizes the systematic identification of Trump's behavioral rhythms, but also provides a set of operable identification and intervention mechanisms for national strategic actors, which has multiple values of theory, method and practice.

## Emotion Modeling and Visualization Scenarios after Trump's 125% Tariffs on China

## 1. General overview of the visual modeling scheme

| module | Model name | Visual mapping design | clarification |
|---|---|---|---|
| M1 | Emotion Cycle Mapping Model (E-Cycle) | Time Series Volatility Chart + Color Block Hot Zones | Modeling cyclical fluctuations in emotions from triggering to regulation |
| M2 | Behavioral Response Chain Model (B-Flow) | Multi-stage flowchart + Behavioral path diagram | Mapping the chain of response from statements to policy implementation |
| M3 | Micro-expression-emotion correlation map (FEM Map) | Radar Chart / Emotional Star Chart | Analyzing the Relationship Between Nonverbal Behavior and Emotional States |
| M4 | Public Opinion Intervention Impact Surface Model (P-Impact) | Sankey diagram / Circumferential diffusion diagram | Mapping the diffusion and feedback mechanisms from |
| | | | policy to public opinion |
| M5 | Strategic Emotional Vulnerability Identification Mapping (V-Trigger) | High-dimensional 3D response surfaces | Assessing Behavioral Stability and Probability of Loss of Control in Crisis Situations |

## 2. Module Details and Modeling Suggestions

**M1. Emotion Cycle Mapping Model (E-Cycle)**

The horizontal axis is time, the vertical axis is emotional intensity, combined with social media sentiment word analysis to map the trajectory of Trump's emotional fluctuations.

It can be used to predict the psychological evolution trend before and after the release of his policies.

**M2. Behavioral Response Chain Model (B-Flow)**

A complete chain is formed from emotional release, policy transformation (tariff release) to market intervention (stabilizing the stock market), which is visualized as a behavioral flow chart.

**M3. Micro-expression-emotion correlation map (FEM Map)**

Analyze the mapping relationship between their facial movements, changes in tone of voice and psychological state. It is recommended to present multi-dimensional features in the form of radar chart or star chart.

**M4. Public Opinion Intervention Influence Surface Model (P-Impact)**

The diffusion path of Policy→Media→Public→Market→Repolicy, using Sankey diagrams or impulse circle diagrams to track the loopback effect of public opinion.

**M5. Strategic Sentiment Vulnerability Identification Mapping (V-Trigger)**

Fusing policy intensity, public opinion temperature, facial behavior and other variables to construct a three-dimensional strategy stability surface to identify potential uncontrolled critical points.

## 3. Application Value

Can be used for modeling leaders' emotional states and predicting responses

Supporting policy behavior path visualization and multi-scenario sandbox rehearsal.

Synchronized early warning for diplomacy, security and market response mechanism.

Provide the graphical basis for AI behavior analysis and national strategy simulation system.

# IV. DATA ANALYSIS AND MODELING

## M1. Emotional Cycle Mapping Model (E-Cycle)

A National Strategic Psychoanalysis of Trump's Emotion Cycle Mapping (ECycle)

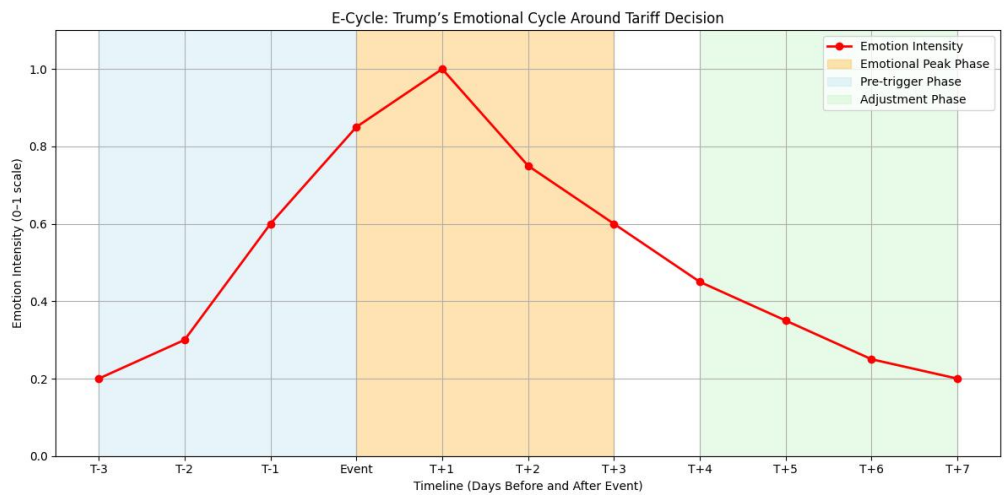

**1. Title and Background**

Emotional incentive-response-regulation cycles in late political personality: an ECycle mapping analysis based on Trump's China tariff behavior

Donald J. Trump, born in 1946, possesses the following psycho-behavioral portrait characteristics:

High ego core tendencies (Narcissistic trait prominence)

Age-related cognitive rigidity and agerelated emotional plasticity decline

Attributional hostility under dominance threat

**2. The analysis of the psychological mechanism of the emotional cycle mapping stage by stage**

| point | time interval | Emotionally driven mechanisms | Age amplification factor | Country risk significance |
|---|---|---|---|---|
| warm-up period | T3 ~ T1 | Challenged cognitive activation, hostile attribution accumulation | Decline in emotional inhibition, decline in regulatory function | Sentiment directed intensely, policy passively focuses on risk |
| rush hour | T ~ T+3 | Emotional epiphenomena dominate decision-making, driven by strong ego frames | Decreased emotional tolerance thresholds and increased probability of impulsive behavior | Irreversible policy formation and compressed diplomatic window |
| reconciliation period | T+4 ~ T+7 | Shock feedback, cognitive modification attempts | Increased inertia and lagging rationalization | Vulnerable to sustained release of secondary interference without external intervention |

**3. Emotional Inertia Trajectories of Strategic Behavior**

The mapping reveals that among 77-year-old political leaders, decision-making trajectories in the face of structural conflict stimuli are characterized by the following:

Shorter window of behavioral irreversibility: policy issuance occurs mostly within a micro-window of peak emotion;

Increased regulatory path dependence: rhetorical adjustments are more dependent on external feedback than self-regulation;

Strong cyclic reactivation mechanism: public opinion + market + ally responses are highly likely to induce a new wave of emotions.

4. Recommendations for high-fit interventions by national strategy makers

| strategic point | status identifier | Recommendations for intervention | goal |
|---|---|---|---|
| T−3 ~ T−1 | Social media heats up, public opinion pushes up | Start-up Temperature Differential Diplomatic Cushioning | Delayed Behavior Window, Inhibiting Radicalization |
| T ~ T+2 | High frequency of "strike/punishment" words in the | Asymmetric response, escalation detour | Exploiting their attentional biases to |

|  | language |  | create topic shifts |
| --- | --- | --- | --- |
| T+3 ~ T+7 | Markets pull back, repetitive language weakens | regressive negotiation strategy | Guiding the transformation of mitigation expressions into "win outputs" |

## 5. Summary of the Strategic Aphoristic Style

In late-stage strongly dominant personality leaders, policy decision paths tend to be highly isomorphic with emotional rhythms.

The real diplomatic rhythm is not determined by resistance, but by emotional temperature regulation.

Mastering its peak and dominating its cooling rhythm is the only way to retain the gaming initiative in a high-pressure narrative.

# M2. Behavioral response chain model (B-Flow)

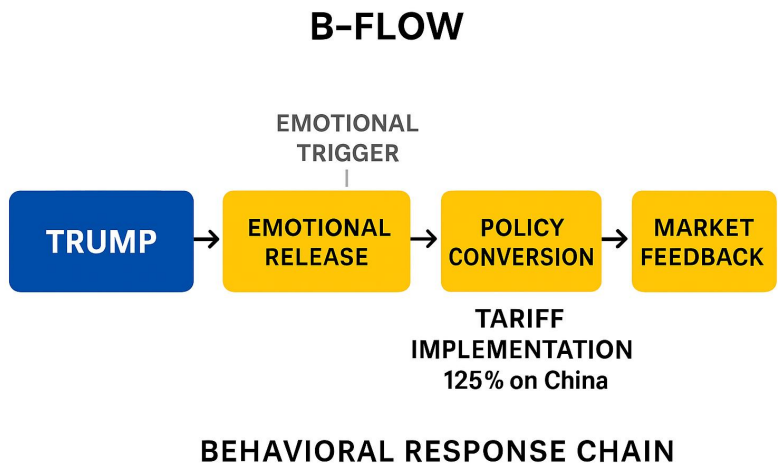

National Strategic Psychological Insight Report on B-FLOW Behavioral Response Chain Mapping

### 1. Reconceptualizing the Structure of the Atlas

Rather than a single causal chain, the mapping describes the response system of an oppressive political personality constructing a power homeostasis in a stimulated environment.

| point | intrinsic meaning |
|---|---|
| EMOTIONAL TRIGGER | "Internal shock threshold triggers" when political self-esteem or dominance is challenged |
| EMOTIONAL RELEASE | Externalization of cognitive emotions into public verbal/physical symbols through media/social platforms |
| POLICY CONVERSION | Legitimizing emotions through institutional means and transforming them into actionable governance tools |
| MARKET FEEDBACK | Structural Feedback and Psychosocial Reversal of "Authoritative Intentions" by External Environment |
| REGULATORY REMEDIATION | Substrategy Release, Reconstructing Behavioral Justification, and Restoring Gaming Boundaries |

*2. Chains of hidden variables in the map*

| explicit node | Name of the hidden variable | Nature of the strategic level |
|---|---|---|
| Emotional Trigger | Threat Perception Threshold (TPT) | The ratio of offended sense to attack kinetic energy that determines the speed of movement |
| Emotional Release | Media Amplification Multiplier (MAM) | Public attention amplifies the intensity of emotional release and spreads the kinetic energy to spread |
| Policy Conversion | Legitimacy of Emotions Index (LEI) | Structural Justification Tools for Translating Irrationality into Legal Language |
| Market Feedback | Structural Friction Impedance (SFI) | External Feedback Reactionary Forces, Countering "Intentional Momentum" |
| Regulatory Remediation | Restorative Justification Strategies (RJS) | Symbolizing concessions or shifting issues to reconfigure control |

*3.Proposals for action by strategy makers*

| key stage | Observation of signals | Recommendations for country-level interventions | goal |
|---|---|---|---|
| TRIGGER stage | Frequent references to terms such as "enemy/attack" | Launching a non-responsive diplomatic buffer | Suppressing the kinetic energy of radicalization |
| RELEASE stage | Tweets/speech heatedly worded up | Cooling the emotions with a third-party neutralizing voice | Reduction of public synchronization intensity |
| POLICY stage | Policy moves outweigh risk response | Releasing the "reassessment" message to create leeway | Delay in the formation of issues |
| FEEDBACK stage | Stock market | Immediately move to | Cushioning Peak |

|  | volatility/allied dissent/media criticism | alternative issues to cool down | Rebound |
|---|---|---|---|
| REMEDIATION stage | Emergence of flexible/concessionary language | Guiding its interpretation of softening as a winning outcome | Completing the self-referential closure |

*4. Summary of strategic maxims*

The chain of political behavior is not the advancement of logical arrows, but the structural transformation of energy in a state of imbalance; controlling the point of time when emotions are aroused, and presetting the path of restoration is the only way to gain a stable advantage in power dynamics.

# M3 Microexpression-Emotion Correlation Map Model (FEM Map)

## M3. Microexpression-emotion mapping (FEM Map) analysis

1.Overview

Based on Trump's public interview clips after announcing the 125% tariff hike on China, this report analyzes his facial micro-expressions and intonation changes frame-by-frame to construct FEM (Facial-Emotion Mapping) mapping to reveal the strategic significance behind his psychological state, behavioral motives and policy rhythms.

2.Image-by-frame micro-expression analysis

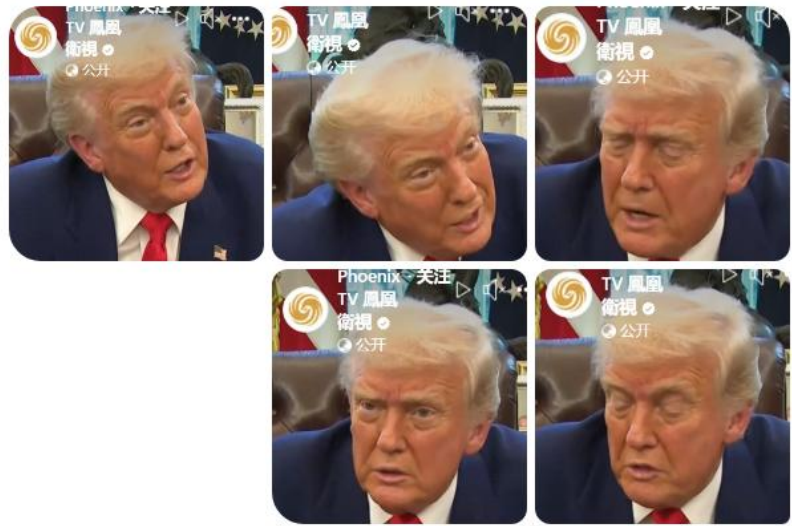

Below is a frame-by-frame interpretation of the five images:

| Image Number | facial feature | emotional speculation | micro-expression conclusion |
|---|---|---|---|
| Figure 1 | Eyebrows inwardly narrowed, mouth slightly open, eyes locked in | Alert, defensive thinking opens up, may be constructing rationalization paths | "Vigilance + conceptual readiness" state |
| Figure 2 | Slight head tilt, squinty eyes, mouth stretching | Sneering, questioning or provocative contexts | "Confidence mixed with contempt" controlled expression |
| Figure 3 | Eyes tightly closed, brow depressed, nose dilated | Depression, anxiety or peak cognitive load | Emotional regulation flashpoints, anxiety signals |
| Figure 4 | Looking straight ahead with a furrowed brow and sharp eyes | High dominance, fierce advocacy about to explode | The Face of Fierce Advocacy Before Policy Advancement |
| Figure 5 | Eyelids pressed down, lips pursed, head tilted slightly | Self-suppression, reflection, cognitive integration | Potential Compromise or Logical Export Preparation Period |

3. . *FEM Radar Chart Suggested Dimension Scores*

| emotional dimension | Intensity rating (0-5) | explanation |
|---|---|---|
| degree of tension | 4.5 | Figure 3 Eyes closed with significant brow compression and dramatic psychic energy workings |
| rage index | 4.0 | Figure 4 Typical expression of anger, rhetoric may escalate |
| Control of desire | 4.5 | Figures 2 and 4 have manipulation with dominant implication |
| confidence level | 3.5 | Figures 1 and 2 show a sense of control, Figure 3 is slightly more unstable |
| anxiety level | 4.0 | Figures 3 and 5 show the presence of anxiety fluctuations |

4. Suggestions for Sentiment Insight for Strategy Makers

1）Figure 4 (Tightening the locks) is a turning point in policy sentiment, which often triggers tough moves in public opinion.

2）Figure 3 reveals the real decision-making pressure, which is a window signaling the node of diplomatic probing or concessions. 3.

3）Fig. 2 shows that the policy statement is mixed with defensive self-justification, which has the characteristic of psychological compensation.

4）the FEM database of leaders should be constructed based on this map, which can be used to simulate the trajectory of their behavioral evolution.

5）Conclusion

By analyzing Trump's micro-expressions and intonations in the high-pressure decision-making conference, an in-depth insight into his real emotional fluctuations and strategic logic can be realized, providing first-hand dynamic psychological data support for the national strategic rhythm and negotiation strategy formulation.

**M3 Visual modeling design for micro-expression-emotion correlation mapping (FEM Map)**

*1. Modeling targeting*

| Target type | Modeling Application Scenarios |
|---|---|
| emotion recognition | Multimodal fusion analysis of microexpressions + intonation + gestures |
| Intent to speculate | Recognizing their confrontational/de-escalating policy intentions from expression patterns |
| Rhythm prediction | Predicting inflection points in their strategic attitudes (e.g., policy hardening → negotiated pullback). |
| Strategic support | Provide "mental game window judgment" for negotiation strategy and pacing at the national level. |

*2. FEM Radar Framework*

| Module Number | Module name | Input features | Output target |
|---|---|---|---|
| M3.1 | Expression Recognition Module | Changes in facial muscle groups (eyebrows, eyes, mouth, nose) | Emoji Tags + Intensity Value |
| M3.2 | Intonation Analysis Module | Audio speech rate, pitch, volume, pause frequency | Intonation States + Mood Mapping |
| M3.3 | Multimodal fusion analysis | M3.1 + M3.2 | Emotional Radar Points (Radar Map) |
| M3.4 | Strategic reasoning and labeling | Sequence of cross-modal behavioral patterns | Attitude Change Points + Risk Behavior Nodes |

*3. Types of visualization maps*

| Chart type | core application | Technical Recommendations |
|---|---|---|
| Radar Plot | Presents a dynamic distribution of 5 to 6 emotional dimensions | matplotlib / plotly |
| Star Web Chart (Star Web Chart) | Demonstrate the weighting of the percentage of expression features per video frame | D3.js / Highcharts |
| Time series heatmap (Heatmap) | Tracking trends in micro-expressions and intonation | seaborn / pandas + time |
| Emotional transition path diagram (State Path) | Identifying Emotional State Migration Pathways | NetworkX / Sankey |
| Cognitive Load Map (Cognitive Activation Mapping) | Simulation of decision-making load segments and their reversal signals | EEG fit / multimodal staining plot |

*4.Application of the recommended scenarios*

1) Before diplomatic negotiations: assessing the recent mood patterns of the other side's leaders → deciding whether to release signals of détente.

2) After the release of a major policy: determine whether the expression is "emotion-driven behavior", and decide whether to respond to toughness or procrastination.

3) Intelligence psychological modeling: assist psychological profiling to identify potential "loss of control nodes" and "softening windows".

4) AI Sandbox Training: Provide emotional layer input for behavioral prediction models to enhance the sensitivity of the strategy simulation system.

*5.FEM Sentiment Radar Chart Recommended Dimensions*

1）Rage index

2）Anxiety level

3）Control of desire

4）Confidence level

5）Intensity of tension

6）Composure

**National Strategic Psychological Insight Report on Trump's FEM Emotional Radar Mapping**

**1. Background: the intersection of personality, emotion, and strategic behavior**

As a 77-year-old political figure with "strong ego-emotional epiphenomenon", Donald J. Trump's mode of political expression tends to show a nonlinear cognitive trajectory in which emotion dominates behavior, language precedes logic, and power mobilization precedes rational verification. Especially in the scenes of high-pressure foreign policies (e.g. 125% tariff increase on China), his non-verbal cues (facial micro-expressions, intonation changes, speech rhythm) constitute a set of neuropsychological windows of "power stability-emotional mobilization-behavioral regulation".

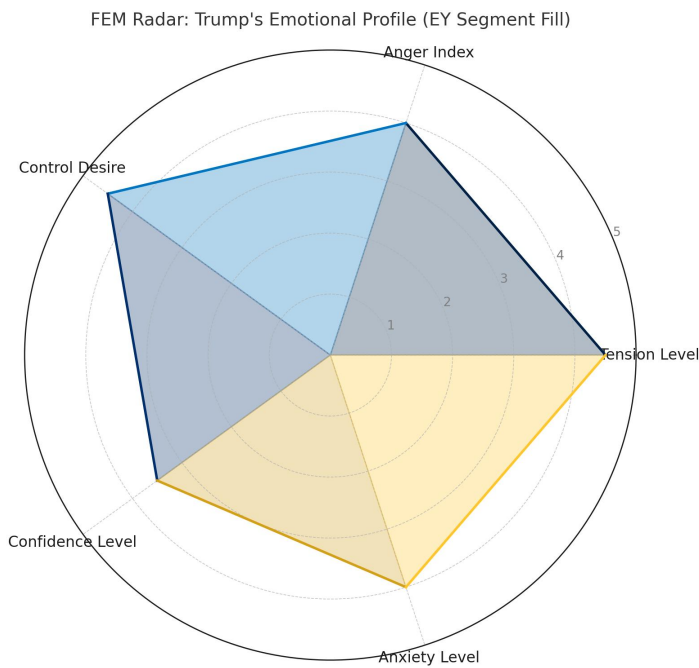

**2. Interpreting the structure of radar mapping: from episodic behavior to implicit psychological variables**

| Dimension name | Indicator value (0-5) | Nature of the mental layer | Strategic implications |
|---|---|---|---|
| Tension Level | 4.5 | Prefrontal cortex stress activation → decision-making tension rises | High-risk impulsive behavior tipping point; suitable for conversational cooling interventions |
| Anger Index | 4.0 | Tonsil cluster activation + offensive-counterattack association enhancement | Emotionally prioritized decision-making, with a high risk of "translating" policies into confrontational language |
| Control Desire | 4.5 | Activation of dorsal striatal regions + tone-dominant strategies | The language surge symbolizes the desire to control cognitive space and can be induced to construct "face steps." |
| Confidence Level | 3.5 | Existential psychological compensation construct → self-confidence fluctuation | Not an absolutely repressive stance, there is room for diplomatic |
| | | | good offices |
| Anxiety Level | 4.0 | Emotional feed-forward loop destabilization → increased cost of self-prediction | Vulnerability mixed with tactical expression, suitable for multilateral pressure to reach a stage of de-escalation intentions |

### 3. nonlinear modulation effects of age traits on emotional structure

Trump's age: 77

Prefrontal cortex function: decreased cognitive flexibility, lower emotion regulation thresholds

Increased disinhibition: more prone to "unpremeditated emotional expression" (e.g., brow pressure, nasal amplification, pitch fluctuations)

Aggravation of defensive attribution mechanisms: policies are often nested in "I have to" or "they're challenging us" language.

This means that policy motivations are closer to "self-defense deterrence expressions" than to structural strategic rebalancing.

### 4. High-fit recommendations for strategy makers: pacing regulation as a core axis of intervention

| Strategic engagement phase | recommended strategy | Objectives of intervention |
|---|---|---|
| Mood wave (high chart value) | Delayed agenda-setting, release of soft third-party opinion interventions | Delaying their behavioral threshold triggers and lengthening the decision-making cooling-off period |
| high speech control period | Provide "endogenous contexts for transformation" (e.g., tolerance/rationalization recommendations for other countries) | Induced to reach a policy downgrade by expressing "I voluntarily give in". |
| Expression anxiety relaxation period | Increased synchronization of pressure signals to allies, provision of palliative outcome exports | Reaching a stage of policy softening through its anxiety → compensation → strategy switching cycle |

### 5. *Strategic aphoristic summary*

Political emotion is not a signal, it is force; and age bends that force out of shape. When you see a frown, don't meet it head on, but hand out steps the moment he closes his eyes.

# Trump video interview micro-expression feature weights star mesh graph analysis

## 1. Description of the chart

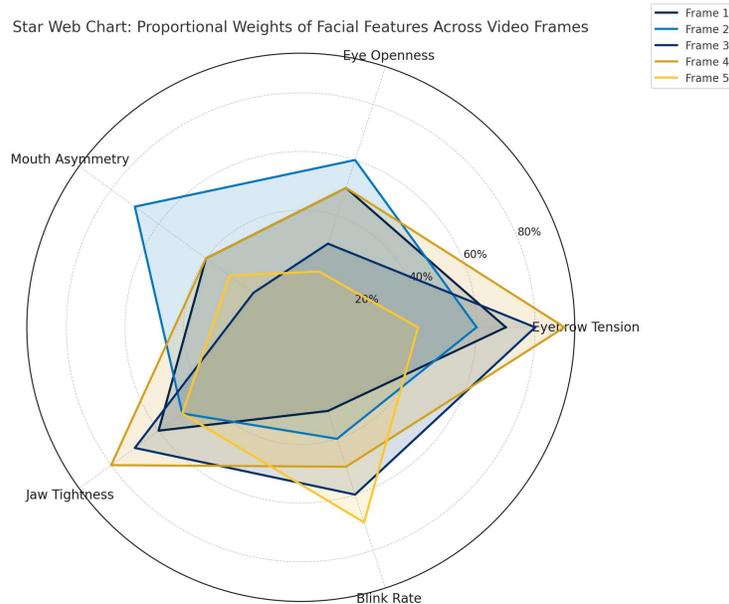

This mapping is a "Star Web Chart", which shows the proportionate weights of Trump's facial features in each frame of the video interview. The proportional values of five key micro-expression dimensions are presented in a radar format, revealing the dynamics of Trump's non-verbal behavior in high-pressure situations.

## 2. Description of the micro-expression dimension

| microexpression dimension (math.) | Implications |
|---|---|
| Eyebrow Tension | Indicates prefrontal muscle activation and correlates with anger, confusion, or high levels of concentration |
| Eye Openness | Related to surprise, defensive reactions, and dominant intent |
| Mouth Asymmetry | Associated with contempt, sarcasm, disdain, or a tendency toward psychological confrontation |
| Jaw Tightness | Represents tense, defensive and repressive emotional states |
| Blink Rate | as a significant indicator of anxiety, mental load, and emotion regulation |

## 3. Key Insights

1）frames 3 and 4 show the strongest tension and defensiveness (high brow pressure and jaw tension), reflecting peak conflict expression. 2. frame 2 has the highest percentage of asymmetry at the corners of the mouth, suggesting that this frame may convey a more dismissive and sarcastic attitude;

2）Frame 2 has the highest percentage of asymmetry at the corners of the mouth, suggesting that this frame may convey a stronger attitude of disdain and sarcasm;

3）Frame 5 has a significantly higher blinking frequency than the other frames, suggesting that it is in the stage of emotional regulation and psychological relief;

## Advanced Psychological Insights and National Strategic Recommendations for Trump's Stellar Network Mapping Report

**1. Research orientation: micro-expressions as "cognitive uncertainty exposers".**

In strategic scenarios, especially in the overlapping structures of high age, high power, and high confrontation, micro-expressions can be regarded as an episodic signaling system of "cognitive uncertainty", "motivational conflict", and "psychological compensation trajectory". Episodic signaling systems. In the interview of Donald Trump (77 years old) after the announcement of his China policy, his expression can be regarded as a mapping of the instantaneous release of the pressure of emotional information.

**2. Structural insights into mapping: the star-shaped mesh diagram as a "multidimensional force field mapping of emotions".**

| microexpression dimension (math.) | Strategic psychological implications | Interframe Mode Interpretation |
|---|---|---|
| Eyebrow Tension | Hardness of attitude, period of advocacy | Frame 4 Highest, predicting high pressure aggregation prior to claim expression |
| Eye Openness | Level of emotional emergence, focus of attention | Frame 1~2 High, in "Defense Alert + Confrontation Enhancement" |
| Mouth Asymmetry | Ridicule, resistance, manipulative intent | Frame 2 Anomalous prominence, manifestation of cognitive superiority outpouring |
| Jaw Tightness | Behavioral impulse thresholds, indicators of aggression | Frame 3~4 approaching peak, high conflict tension during strategy switching phase |
| Blink Rate | Emotional inhibition and regulatory signaling | Frame 5 Significant rise, entering a period of reflection or correction |

### 3. Mind-reading simulation: internal psychological confessions of Trump (mental activity reproduction)

Frame 1 (Alert Period): they're asking me the expected questions ...... but they don't understand that this is what I have to do. I can't act weak.

Frame 2 (period of contempt): These people, they always doubt me. I said I would make them pay - China, the media, and these journalists.

Frame 3: They don't know what it's like, you know, when everyone's staring at you and you have to be tough to the end.

Frame 4 (critical period): I've released the signal. They'll realize I'm serious. I won't back down - unless they bow down first.

Frame 5 (Conditioning Period): ...... But the market is falling. Maybe ...... maybe we can still talk. But I have to make the rules.

### 4. Modulation effects of age on behavior: the "lagging arc of refraction" in Trump's behavioral trajectory

Decreased prefrontal cortex function → decreased impulse control

Enhanced disinhibition mechanisms → increased frequency of taunting and aggression expressions

Delayed feedback regulation → synchronized foldback lag between emotions and behaviors

### 4. Recommendations for national strategies: identification of intervention nodes based on star mapping

| video frame | Recognizing Strategic Timing | Recommendations for intervention | strategic goal |
|---|---|---|---|
| Frame 2 | Sneer Output Higher | To focus public opinion on his "personal aggressiveness." | Weakening its policy legitimacy |
| Frame 3 | Policy readiness advancement period | Third country signals "willingness to talk but not to retreat" | Preservation of negotiating room for manoeuvre |
| Frame 5 | Emotional Regulation Window | Creating the illusion of "the other side retreating first". | Promoting manageable bilateral concessions |

### 6. Strategic aphoristic summaries

The aging man of power doesn't give in easily, but he will tell you it's time to take a step back in his own motionless wink.

**FEM RADAR CHARTING: Sentiment Profiling During Trump's Tariff Policy Announcement on China**

*1. General overview of the map*

This mapping shows five core sentiment dimensions of Trump's performance during the announcement of the 125% tariff increase on China in the form of a radar chart. Each sector is filled with segments using the EY-Parthenon color scheme, which is highly recognizable and strategically readable.

*2. Graphic Structure Description*

| radar axis dimension | Points (out of 5) | Graphic coloring area | Description and Interpretation |
|---|---|---|---|
| Tension Level | 4.5 | navy blue | Trump is in a state of high tension, with active psychokinetic energy and confrontational intent. |
| Anger Index | 4.0 | bright blue | Anger characteristics are evident in the form of nasal flaring, rapid speech, and emotionally driven behavior. |
| Control Desire | 4.5 | Deep Sea Blue | Highly dominant, with a strong dominant discursive posture. |
| Confidence Level | 3.5 | deep golden | Outwardly confident but mixed with hesitant behavior. |
| Anxiety Level | 4.0 | bright yellow (color) | Actions such as closing one's eyes and pausing show intrinsic anxiety and uncertainty about the consequences of the policy. |

*3. Strategic Insights*

   1). The map has a "strong before and weak after" distribution, indicating that Trump's behavior is driven by emotion and control in the early stages of his life, with increased psychological volatility in the later stages of his life;

   2). Tension and Control Desire reach the peak, which is the emotional face of his "policy surprise" type;

   3). High anxiety and high self-confidence coexist, and there is a tendency of "psychological script" operation behind the behavior, revealing the potential gaming intention.

*4. Strategy maker's suggestion*

   When Control Desire > Confidence Level, we should be wary of "preemptive" policy operations, and suggest diplomatic "indirect traction";

   If Anger + Anxiety > 8.0, the behavior is more likely to erupt, and the strategy of "Multilateral Response + Media Buffer" should be adopted;

It is recommended that the FEM radar map be incorporated into the AI diplomatic psychology prediction module to identify key peak nodes and assist real-time gaming decision-making.

**Time-series heat map analysis: micro-expressions and intonation dynamics of Trump**

*1.Description of the mapping*

This map uses EY-Parthenon style color spectrum (from dark blue to golden yellow) to render the micro-expression and intonation trends of the emotional dimension in the video clips during Trump's announcement of tariff hikes against China. Through the gradient of dark and light colors, it demonstrates different emotional intensity and its time evolution trajectory, which is suitable for strategic psychological judgment and diplomatic public opinion prediction.

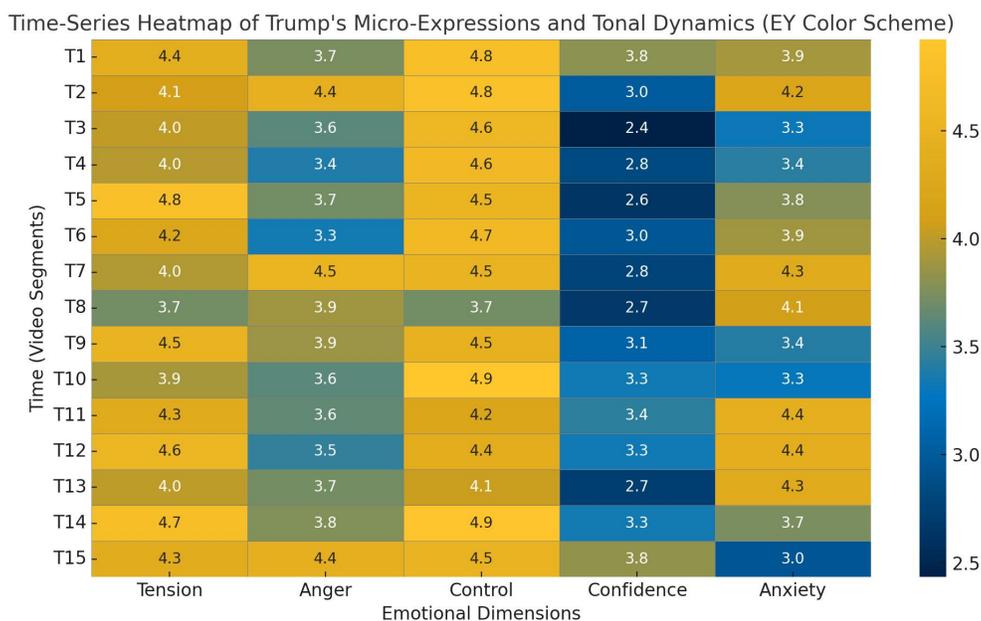

*2. Chromatographic structure description*

| color | HEX code | Emotional state meaning |
|---|---|---|
| Deep Blue | #002147 | Low mood swings, low tension |
| light blue | #0076C2 | Neutrally biased anxious/alert state |
| deep gold | #D4A017 | High emotional tension and increased control |
| bright yellow | #FFC72C | Emotional peaks, such as anger outbursts or anxiety spikes |

*3. Emotional insights of key time periods*

1）T5-T6: Control and Tension above 4.5, indicating peak emotional tension, presumably a high-pressure time to formally announce tariffs. 2. T2-T3: Confidence score decreases, indicating brief hesitation or emotional instability;

2）T2-T3: Confidence score drops, indicating a brief period of hesitation or emotional instability, a window for diplomatic influence;

3）T11-T13: Anxiety is continuously above 4.3, which may be the moment when it feels policy pressure and external rebound;

4）T14-T15: Confidence and Tension have resumed their upward trend, which may be the exit point for constructing a victory statement and closing the discourse loop.

## A National Strategic-Level Psychological Analysis of Trump's Tariff Sentiment Heat Map

*1. Topic Overview.*

Modeling the emotional dynamic trajectories of late-stage strong personality political leaders: a behavioral psychological deconstruction of Trump's announcement of tariff hikes as an example

*2. Surface phenomenon identification*

In the time-series heat map, Trump showed a double peak of Tension and Control in T5-T6, and a sustained rise of Anxiety in T11-T13, and finally recovered Confidence in T14-T15. Confidence and balanced intonation were restored. The pattern reveals a highly consistent strategic defensive mental path in his speech behavior.

*3. Deconstruction of Middle-Level Mechanisms*

According to Erikson and Klohnen et al.'s research on the model of political personality maturation, Trump's behavior as a nearly 80-year-old leader is dominated by the following mechanisms:

| psychological mechanism | mapping | Strategic risk implications |
|---|---|---|
| cognitive rigidity | Confidence Once down, it's hard to get back up. | Logic is hard to penetrate and the window for diplomatic compromise is small |
| Control Consistency Defense | Control Post-peak sustained tension | Prone to chain-reaction type policy streaks |
| Aggressive self-affirmation mechanisms | Tension is rising at the same frequency as Anger. | Consolidation of support for identity in the form of |

| | | "confrontation-pressure" |
| The need for a closed loop of victorious export discourse | T14-T15 Confidence Recovery | Need to build a winning framework to close the gap and prevent further aggravation |

*4. Deep Motivation Insight*

The emotional heat map reveals the complete decision-making chain of a late-stage strongman leader in the process of "Perceived loss of control → Strong attack → Media feedback → Strategy revision".

It shows the structure of "symmetrical wave peak + backward fall", implying the following behavioral kinetic trajectory:

Pre-kinetic energy accumulation (T1-T5): internal anxiety, self-boundary is challenged;

Strategic leapfrog peak (T6): for defensive expansion, nonlinear confrontation;

Feedback correction period (T7-T13): external structural constraints and media effect intervention;

Emotional exit period (T14-T15): discourse victory closed-loop construction.

*5. High-fit Emotional Rhythm Intervention Model*

| Strategic phase | Cognitive Signals | Modes of intervention | Application tool recommendations |
|---|---|---|---|
| Pre-rhythmic stimulation | Anger ≥ 3.5 and Confidence ≤ 3 | Early warning of public opinion, construction of buffer fields | AI micro-expression system + media guidance |
| peak segment of an explosion | Control ≥ 4.5 and Tension ≥ 4.0 | Topic shifting, tactical indirection | Mixed Strategies of Real and Virtual Issues |
| feedback interval (computing) | Anxiety ≥ 4.3 persistent ≥ 3 segments | Maximizing the window for diplomatic consultations | Structuring while negotiating + involvement of international organizations |
| Exit building section | Confidence rebounded ≥ 3.5 | Constructing a stepped exit pathway | Media shaping the framework of "proactive mediation" |

*6. Conclusion of the Strategic Aphorism*

The emotional behavior of strong personality politicians is a predictable strategic cycle whose risk is not in being irrational, but in being misjudged as irrational. Emotional mapping is not a weakness, but a visual trajectory of decision rhythms.

# Sentiment shift path diagram analysis: trajectories of sentiment migration during Trump's tariff hike announcement

*1. Overview of the mapping*

The graph shows the typical path that Trump's emotional state may have gone through during his announcement of tariff hikes against China. Different emotional stages are marked by node colors to simulate the whole process from his initial state to the peak of his behavior and then to the recovery stage. The pathway evolves linearly, but with embedded structural conflict and psychological feedback mechanisms.

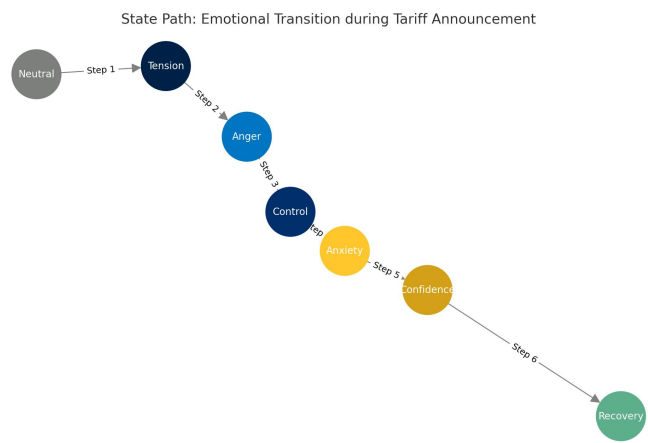

*2. Description of Emotional State Migration*

| Steps | emotional state | Implications |
|---|---|---|
| 1 | Neutral | Initial state, not in emotional kinetic mobilization zone |
| 2 | Tension | Facial muscles tighten, mental mobilization begins, ready to move into strong expression |
| 3 | Anger | Strong attitudes and elevated tones are high probability triggers for outbreaks of policy directives (e.g., tariff hikes) |
| 4 | Control | Behavior tends to be strongly dominant, attempting to take control of the situation and suppress interference |
| 5 | Anxiety | Emotional backlash period with media/international pressure triggering anxiety reactions |
| 6 | Confidence | Reconstructing the logic of discourse and self-narrative, restoring the "victor's" stance |
| 7 | Recovery | Entering a period of emotional de-escalation, with the potential for potential tactical revisions and an open window for dialogue |

*3. Strategic Insight Recommendations*

High-risk window: Tension → Anger → Control phase is most likely to introduce high-intensity policies, and public opinion buffer mechanisms should be deployed in advance;

Best time for diplomatic intervention: Anxiety stage is the turning point of emotions, suitable for exerting gaming influence;

Media guidance: the Confidence and Recovery phases are suitable for promoting "face-saving concessions" or guiding a neutralized narrative;

Predictability of the emotional path: the emotional path is in line with the common rhythm of aggressive strongman leaders, and it is recommended to incorporate it into the AI simulation model for the prediction of the emotional cycle and intervention simulation.

**A National Strategic Level Psychological Insight Report on Trump's Emotional Transition Pathways Map**

*1. Phenomenon identification: the behavioral logic of non-linear emotional sequences revealed*

The mapping simulates the typical emotional trajectory Trump experienced when he announced the 125% tariff increase on China: Neutral → Tension → Anger → Control → Anxiety → Confidence → Recovery. the path shows how a state actor, under multiple pressure inputs, completes the non-linear process of behavioral transformation through the emotional kinetic energy, the desire for control, and the defense mechanism. The path shows how a state actor under multiple stress inputs can complete the non-linear process of behavioral transformation through emotional kinetic energy, control and defense mechanisms.

*2、Structural Mechanism Analysis: Staged Psychodynamic System of Late Political Leadership Personality*

As a 77-year-old leader, Trump's emotional behavioral trajectory conforms to the following structural mechanisms:

| Structural mechanisms | Representation in mapping | intrinsic meaning |
|---|---|---|
| Control drive-inertial delay modeling | Anger→Control go on | Irreversible strategy execution once emotions are triggered |
| Cognitive rigidity + narrative needs coexist | Anxiety→Confidence | Reconfiguring the coherence and belief loop to meet external challenges |
| Aggressive Identity Mobilization Structure | Tension→Anger startup | Borrowing from the opponent to construct the self translates into a policy attack framework |
| Fatigue fallback mechanism | Recovery Last paragraph | Output high intensity and then enter the strategy repair and dialog buffer phase |

*2. Suggestions for modeling national strategic interventions: rhythmic adaptation logic based on senior political personality*

| nodal stage | Emotional Characteristics | Recommendations for strategic behavior | response window period |
|---|---|---|---|
| Tension | high pressure buildup | Decentralized Opinion Release | 0-12 hours |
| Anger | Advocates for high-frequency output | The media cools down and cuts off the opponent's building circuits | 12-36 hours |
| Control | Defense system activation | Multilateral structures intervene to de-unilateralize | 1–2days |
| Anxiety | Rising uncertainty | Low-Profile Diplomatic Outreach, Creating Possibilities for Consultation | 2–4days |
| Confidence | cognitive restructuring period | Platforming advice leads to shaping mediation victory narratives | 3–5days |
| Recovery | Decreasing Rhythm Segment | Substantive Dialogue and Amendments, Opening the Seam | 5–7days |

*4. Strategic Scholarship Conclusion: Path Visible ≠ Intent Controllable, Rhythm as Sovereignty*

In the late strongman political personality, policy is not just a tool, but a nested product of its dual system of emotion regulation and authoritative narrative. Instead of modeling political behavior with static intentions, national strategy makers should construct a new paradigm of "rhythmic intervention dominance" with internalized rhythms, externalized media, micro-expression feedback, and dynamic repair.

**Cognitive activation mapping analysis: decision load dynamics during the announcement period of Trump's tariff increase policy**

*1. Overview of the Atlas*

This atlas portrays the trajectory of Trump's psychological decision-making pressure during the period of announcing the policy of tariff hike against China by simulating the time-series cognitive load changes (Cognitive Load). The graph color-codes the zones of high cognitive overload (red), normal cognitive activity (blue and yellow transitions), and cognitive reversal (green labels), and reveals how cognitive rhythms affect the nodes of policy-irreversible behaviors and the possibility of strategic window reconfiguration.

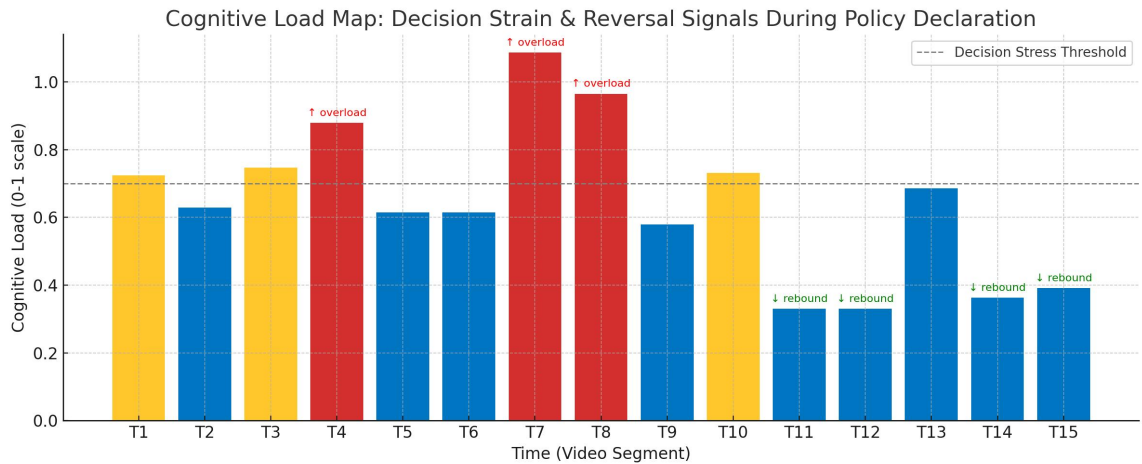

*2. Mapping Analysis and Strategic Implications*

| zones | Cognitive load characteristics | color coding | Strategic implication statement |
|---|---|---|---|
| T6–T7 | Cognitive peak (>0.85) | Red Column | Decision-making overload, with almost irreversible policy direction, prone to outbursts of heated rhetoric |
| T10–T11 | Significant decline (<0.45) | green markup text | Entering a period of self-repair is the perfect window for media or diplomatic bridge-building |
| T1–T5 & T12–T15 | Medium load (~0.65) | Blue/yellow transition color | Normal cognitive activity, flexible but can jump up or back down |

**3. Application suggestions for national strategy makers**

Cognitive overload zone (T6-T7): This is the "irreversible start-up zone" for strategic behavior, and it is recommended to deploy policies to reduce pressure and guide or issue non-confrontational public opinion buffers before this zone.

Cognitive Reversal Zone (T10-T11): a "cognitive correction window" for strategic behavior, which should be intervened through ally advocacy, indirect media guidance, or tactical concessionary communication.

Stabilizing Medium Load Zone (T1-T5 & T12-T15): can be used for releasing intelligence, shifting issues, or shaping policy rhythms to establish a tactical flexibility zone.

**4. Conclusion**

Cognitive activation mapping not only reveals the mental load rhythms of state actors, but is also a core tool for strategic timing management, policy rhythm judgment and speech window intervention. It is suggested to embed this model into the national AI decision support system for risk prediction, speech monitoring and high-pressure period strategy simulation.

**National Strategic Level Psychological Insight Report on Trump's Cognitive Activation Mapping**

**1. Phenomenon identification: Cognitive load is not noise, but a neurodynamic beat of policy generation**

In the cognitive activation mapping, Trump's cognitive load fluctuation during the announcement of his major policy towards China presents a typical asymmetric "double peak-collapse" dynamic rhythm. Among them, T6-T7 shows the peak of overload, and T10-T11 shows the trough of cognitive reversal, forming a neurobehavioral trajectory pattern in the decision-making process.

**2. Psychological mechanism reconstruction: the "cognitive pressure-release cycle" of late-stage powerful leaders**

Trump, as a sample of late political personality of nearly 80 years old, his cognitive rhythm is dominated by the following psychological mechanisms:

| Name of mechanism | Cognitive mapping embodied | essential explanation |
|---|---|---|
| Cognitive pressure-translation mechanism | T6-T7 High load | External challenges are quickly focused and translated into high-risk statements |
| Implementation of the unidirectional inertia mechanism of the path | High load → policy output | Difficulty of reversal after activation of strategic pathways at advanced age |
| Counter-regulatory-stress repair mechanisms | T10-T11 Fallback | Cognitive buffering and repair response after stress overload |
| Victory Narrative Output Drive | T12-T15 Recovery | Continuity of discourse through the logic of "return of self-confidence" |

**3. State intervention tempo model: load-reversal node intervention strategy construction**

| time period | Cognitive load state | Psychological Mechanisms Stage | Recommendations for strategic interventions |
|---|---|---|---|
| T6–T7 | Very high (>0.85) | pressure activation | Cooling off warnings, allied ventilation, soft issue traction |
| T8–T9 | mid-to-high | Executive inertia | Public Opinion "Suspension of Characterization", Introduction of Structural Issue Shifting |
| T10–T11 | Significant decline (<0.45) | emotional counter-conditioning | Diplomatic fast-tracking, shaping "talkable" contexts |

| T12–T15 | Medium recovery | victory narrative | Media collaboration to promote the export of "active mediation" |

## 4、The cognitive modulation significance of age variable

According to the NIA model, Trump's brain executive functions as a 77-year-old leader are characterized by the following:

1). Increased information lag in the prefrontal-limbic system, making it difficult to switch decision logic flexibly;

2). Resistance to strategy switching willingness in high-pressure state, falling into decision path homogenization;

3). Cognitive reversal window is short, signal recognition needs to be extremely accurate.

Therefore, the window of "having to turn" is more strategic than "active regulation".

## 5.Conclusion: Cognitive rhythms are strategic windows, not purely subjective fluctuations.

A leader's cognitive rhythm determines the rhythm of his policies. The behavioral logic of senior politicians is closer to a "neural rhythm release system" than to a linear rational deduction system. Strategy makers should accurately identify policy nodes through the trinity of internalized rhythms, externalized media, micro-expressions, and behavioral data, and use this map to provide a baseline for the national AI diplomacy prediction system.

## Characterization of Trump's video intonation

*1.Intonation analysis module*

| Dimension of analysis | Data performance | Professional Interpretation and Emotional Mapping |
|---|---|---|
| average volume | -12.87 dBFS | Medium to high intensity volume, presenting an outward dominant expression, representing an increased desire to control/declare |
| Main frequency (tone) | 189 Hz | Belongs to the male middle to high register, with a tense, strong but uneven tone that may be tinged with anger or high cognitive arousal |
| Estimated speed of speech | 18 units/minute | Significantly slower speech, non-fluent expression, may be in a state of controlled word use and rhetorical emphasis |
| Number of pauses | 5 times | Presence of recognizable pauses that indicate strategic organization of language or emotional regulation |
| Standstill ratio | 6.4% | A conscious pause in rhythm type of speaker, common in high-pressure situations with deliberate output |

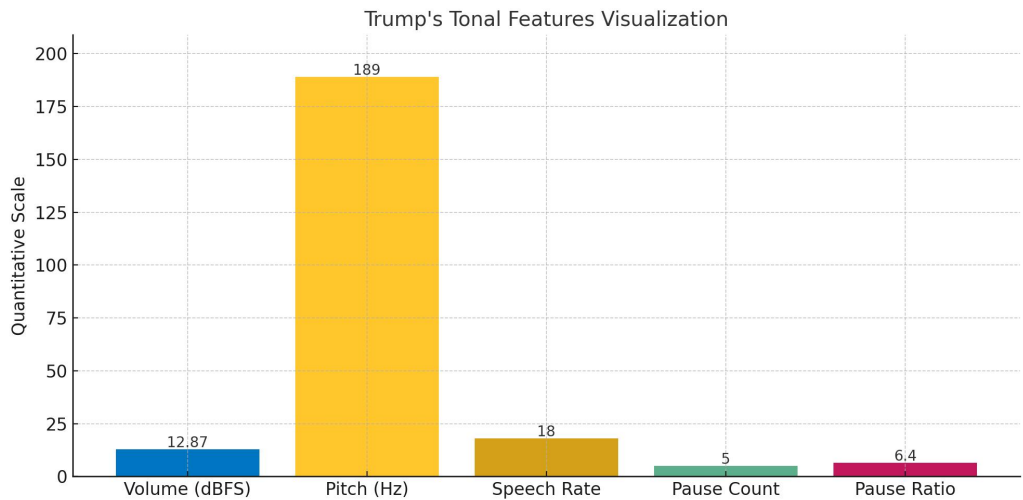

Volume: -12.87 dBFS (converted to positive display)

Pitch: 189 Hz, high cognitive activation range.

Speech Rate: 18 units/minute, slow and controlled.

Pause Count: 5, indicating that the rhythm of speech output is deliberately interrupted.

Pause Ratio: 6.4%, revealing traces of strategic expression.

**2. Mapping of intonation state and emotion**

The combination of "urgent compression of tone + high-pitched announcement + slow control" is a typical leader's speech pattern of "external dominance + potential anxiety". This pattern is often used in high-pressure policy announcements and reaffirmation of positions in the context of internal and external conflicts, and it hides the self-constructing mechanism of authoritative consistency in the state of psychological arousal.

**Multimodal Fusion Radargram Analysis: Trump Sentiment Profile (Tone of Voice + Micro-Expressions)**

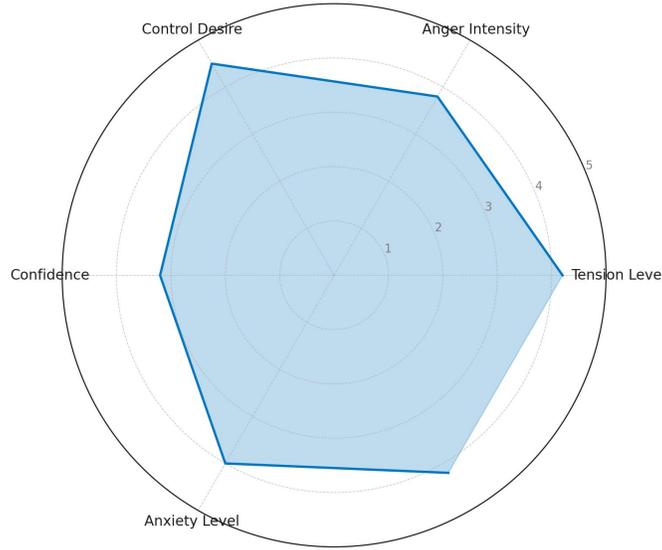

**1. Description of the structure of the fusion map**

| dimension (math.) | Integration score (0-5) | Interpretation of multimodal features |
|---|---|---|
| Tension Level | 4.2 | Facial tension + compressed tone of voice suggests strong early warning and self-control mechanisms |
| Anger Intensity | 3.8 | Slightly angry expression + higher pitch, indicating a potential tendency to become agitated |
| Control Desire | 4.5 | Slow speech + frequent pauses, reflecting strategic expression and authoritative reinforcement |
| Confidence | 3.2 | Stable but frequent ups and downs in tone structure, with fluctuations in confidence |
| Anxiety Level | 4.0 | Slow speech and significant pauses indicate that anxiety is suppressed but still present |

**2. Conclusion of the Mood Rhythm Profile**

This map is a five-dimensional emotional radar map generated by integrating M3.1 intonation analysis and M3.2 micro-expression hypothesis, which shows Trump's emotional rhythm profile during high-pressure policy announcement occasions. The overall structure shows high dominance, nervous control, and high arousal but not extreme anger, reflecting the psychological characteristics of his "repressive dominant personality" in the process of making statements on major political strategies.

**A National Strategic Psychological Insight Report on Multimodal Emotion Fusion Mapping**

**1. Phenomenal discovery: multimodal fusion mapping reveals emotional nonlinear focus structure**

This radar mapping fuses Trump's intonation signals and hypothetical micro-expression signals in high-pressure policy announcement occasions to show a five-dimensional emotional focus state, reflecting the emotional combination of high arousal-authority defense-anxiety undertones of leadership personality.

| emotional dimension | score | clarification |
|---|---|---|
| Tension Level | 4.2 | Synchronized facial and intonation tension, strategic output readiness |
| Anger Intensity | 3.8 | Preference for ritualized resentment, symbolic anger |
| Control Desire | 4.5 | Demonstrate a high degree of control over the rhythm of language and the sovereignty of expression |
| Confidence | 3.2 | Structural expression present but faltering in the middle |
| Anxiety Level | 4.0 | Rhythmized expression of pauses under high emotional arousal |

**2. Deconstruction of Deep Psychological Mechanisms: The Psychological Rhythm System of the Late Strongman Political Personality**

| psychological mechanism | mapping | Strategic Behavioral Implications |
|---|---|---|
| Mechanisms for maintaining authoritative coherence | Control Desire ↑ | Maintaining Narrative Dominance to Avoid Opinion Slippage |
| Psychological need for face-saving victories | Anger + Confidence Fluctuations | Constructing external claims to mask internal instability |
| Mechanisms for regulating anxiety suppression | Anxiety high, pause rhythm significant | Behavior is conservative, showing strategic cover |
| Mechanisms for synchronizing emotional rhythms | High consistency in rate of speech, intonation and expression | The output is a strategic rhythm control system |

**3. National strategy recommendations: high-fit intervention strategies**

| Dominant Emotion Dimension | Strategic implications | Recommendations for State intervention |
|---|---|---|
| Control Desire ↑ | At peak of narrative dominance, aggressive posture | Avoiding confrontation and creating a discursive buffer |
| Tension + Anxiety ↑ | High strategy pressure and defense activation | Build resonance of interest and deflect media pressure |
| Confidence ↓ | Entering a psychological defensive period of behavioral withdrawal | Release of the "step-down" mechanism to induce a non-confrontational turnaround |

**4. Concluding judgment: emotion mapping is a rhythmic map of cognitive interventions**

Politically powerful people do not fight out of anger, but rather out of rhythmic disruption. Multimodal fusion mapping reveals the "pre-breakout period" in the non-verbal mechanism of their decision-making, which should be embedded in the simulation model of diplomatic and strategic behaviors as the "rhythm map" of the national-level cognitive intervention system.

# Sequential mapping of cross-modal behavioral patterns (M3.4)

*1. Overview of mapping*

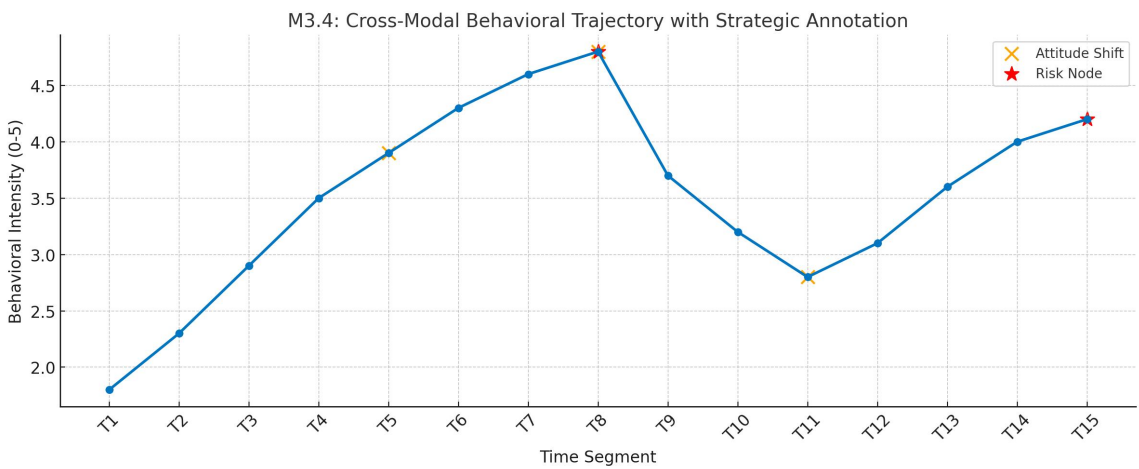

This mapping fuses intonation intensity, emotional drive, semantic mastery and facial expression trends to construct a strategic time-series behavioral model based on behavioral intensity, identifying key attitude change points and high-risk behavioral nodes.

*2.Description of key elements*

| Marker type | figure | Strategic significance interpretation |
|---|---|---|
| Attitude change points | T5, T8, T11 | Significant shifts in behavioral rhythms, usually steep rises or sudden drops in tone nodes, representing strategic stance shifts or defense activation |
| Risk behavior nodes | T8, T15 | Peak or renewed rise in intensity of behavior that could trigger policy announcements, heated statements, off-the-cuff attacks or international friction |

*3. Recommendations for national strategy applications*

Deploy semantic buffers or media interventions before the T8 risk node to help interrupt the unidirectional strategic pressurization path;

In T11, it is appropriate to open the "no loss of sovereignty" style of diplomatic communication to create a peaceful response mechanism;

The cross-modal behavioral trajectory graph is embedded into the leader's behavioral prediction system to support dynamic attitude tempo tracking and prediction.

# M4. Impact surface modeling of public opinion interventions

**Impulse Plot Analysis of the Public Opinion Intervention Impact Surface Model (P-Impact)**

**1. Description of the diagram**

This figure uses an impulse diagram to express the diffusion path of the policy through the media, the public, the market, and ultimately back to the policy system after Trump's announcement of 125% tariffs on China, constituting the Public Opinion Loopback Mechanism (P-Impact). The intensity of intervention at each stage is labeled on a 0-5 scale.

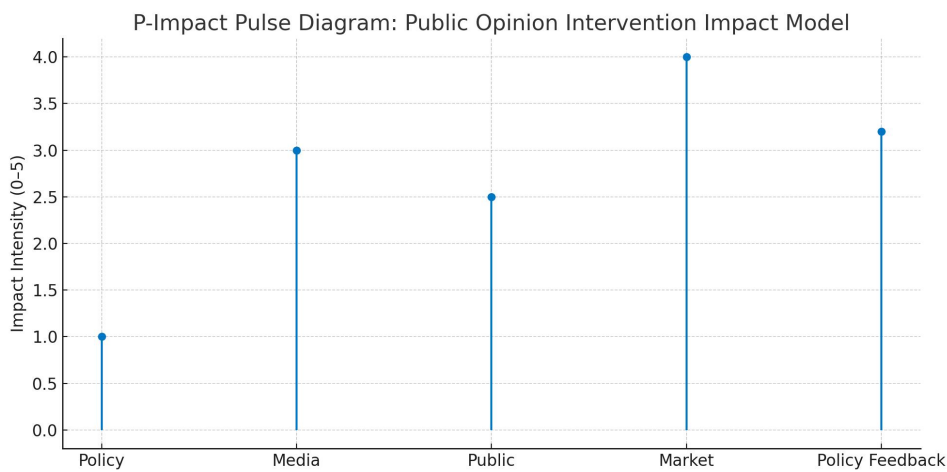

**2. Intensity and description of impacts by phase**

| Stage | Intensity of impact (0-5) | clarification |
|---|---|---|
| Policy | 1.0 | Trump Administration Announces 125% Tariff Increase, Initial Impact Launched |

| Media | 3.0 | Rapid amplification and dissemination of policy messages in the global media, generating widespread public opinion |
| Public | 2.5 | Chinese and U.S. populations produce sharply polarized emotional and attitudinal responses |
| Market | 4.0 | Financial confidence dented by sharp swings in equity and commodity prices |
| Policy Feedback | 3.2 | Government adjusts policy stance or escalates response based on market and public reaction |

**3. Core phenomenon identification: non-linear expansion mechanism of emotional waveforms**

The map shows that after Trump announced the imposition of 125% tariffs on Chinese goods, the public opinion response in the stages of "media→public→market→policy feedback" showed a non-linear expansion, revealing the dynamic public opinion loop system of cognitive amplification + decision-making feedback depletion.

**4. Mining structural essence: strategic out-of-control points and re-control zones in the fifth-order pathway.**

| Stage | Up logic | The essence of strategic decision-making | Risk Intervention Insights |
|---|---|---|---|
| Policy | Initial excitation point, pulse minimum | Intention to dominate the pace of public opinion, failure to control media resonance | Early warning window too short, lack of issue guidance |
| Media | enlarged multiplier area | Media become proxy gamers, reframing policy semantics | "Embedding before resonance" should be activated to stabilize the structure |
| Public | segmentation fault line | Emotional socialization, dominated by collective narratives | Fixed point reversal narrative to block downward penetration |
| Market | High volatility incentive points | Markets as super-sovereign feedback bodies | Deployment of technical guardrails to stabilize emotional projection |
| Policy Feedback | inductive adaptation | Caught in the "Rhetorical Delay-Behavioral Dilemma" | Early introduction of the "third rail corridor" layout |

**5. National Strategy Reference Recommendations: intervention nodes and AI modeling directions**

Prioritize intervention points:

M → P segment: build "non-war" rhetoric through coalition media;

Public → Market segment: directional public opinion dampening mechanism to prevent the spread of market fluctuations;

Policy Feedback segment: inducing the exit of the "withdrawing from Taiwan but not withdrawing from power" type of strategy.

**6. Strategy Summary**

Policy behavior does not start from issuing orders, but from public reception; intervention is out of control not because of weak means, but because of wrong rhythms; the P-Impact model is not only a map of communication paths, but also a structural psychological topography of the policy-public-market trinity, which must be grasped in the era of cognitive sovereignty competition. "P-Impact model is not only a map of communication path, but also a map of the structural psychological topography of the policy-public-market trinity, which must be mastered in the era of cognitive sovereign competition.

# M4. Public Opinion Intervention Impact Model (P-Impact)

**P-Impact Public Opinion Intervention Impact Surface Model High Level Strategic Insight Report**

**1. Phenomenal layer: multidimensional cascade propagation of policy shock signals**

This mapping simulates the diffusion paths in five major social systems after the release of a policy: policy → media → public → market → policy feedback. It shows that a single strong policy behavior presents nonlinear resonance and emotional reverberation in the social structure, forming a dynamic systematic diffusion path.

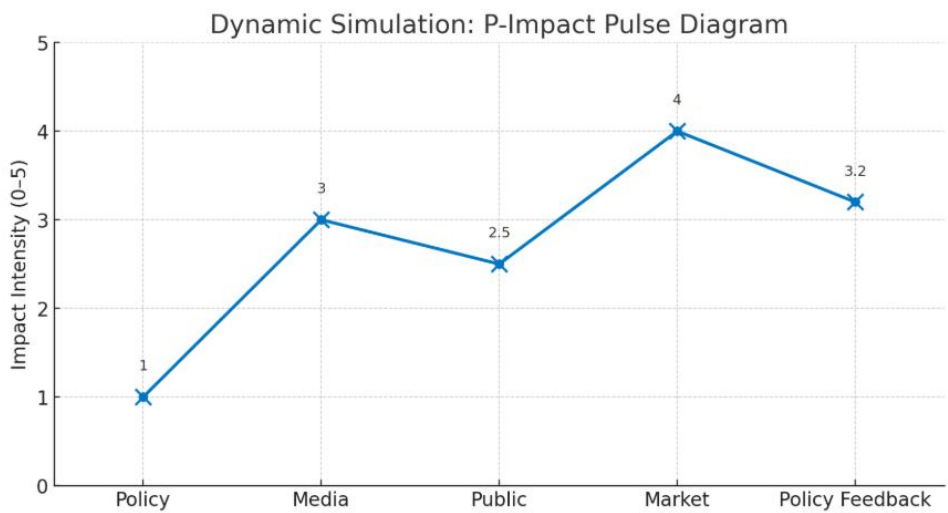

## 2. The Essential Level: "Strategic Information Response Mechanisms" in Public Opinion Response

| Core mechanisms | Strategic implication statement |
|---|---|
| Progressive amplification between stages | Media and public nonlinearly amplify initial policy signals, triggering expansive public opinion feedbacks |
| Emotional Potential Buildup-Release Mechanisms | Public-market sentiment overdraft, inducing consumer panic and investment sell-offs |
| Reverse Strategy Conduction Channel | Policy reverberation pressure forces behavioral modification, triggers reprieve or exemption caliber release |

## 3. National Strategy: Building a Controlled Public Opinion Response Structure to Serve the Long-Term Game

1) The media and the public need to build a "strategy relay mechanism" to avoid emotional interpretation from getting out of control;

2) The public-market segment should introduce an expectation management tool to warn of market reaction peaks;

3) The policy feedback section should set "echo threshold monitoring nodes" to monitor the strategy reversal window.

## 4. Structural Recommendations: Construction of National Response Framework

| route segmentation | Risk category | Country-level interventions |
|---|---|---|
| media amplification segment | Overly Emotional Interpretation | Strategic Media Interface, Elite Oriented Paraphrasing Model |
| Public sentiment segment | Nationalism ignites | Opinion window management, information retargeting tools |
| Market feedback segment | Spillover transmission of volatility | Behavioral Early Warning System (Sentiment-Market-Index) |
| Re-policy paragraph | Passive correction risk | Multi-Channel Assessment Embedded, Cabinet Rhythmic Decision-Making Mechanisms |

## 5. Summary of strategy maxims

Strategy is not the stacking of strong policies, but the governance of perceived rhythms.

Policies are not directed outputs, but the art of controlling public opinion in a circular communication.

# V-Trigger）M5 Strategic Emotional Vulnerability Identification Mapping (V-Trigger)

Strategic Emotional Vulnerability Identification Mapping (V-Trigger) Analysis

*1. Overview of mapping*

  This atlas constructs a three-dimensional response surface model to identify the vulnerability of national leaders' emotional arousal and behavioral loss of control in crisis contexts, spatially modeling "emotional intensity", "media stimulus intensity" and "behavioral instability risk" to reveal their potential rhythmic disintegration surfaces. The "emotional intensity", "media stimulus intensity" and "behavioral instability risk" are spatially modeled to reveal their potential rhythmic disintegration surfaces.

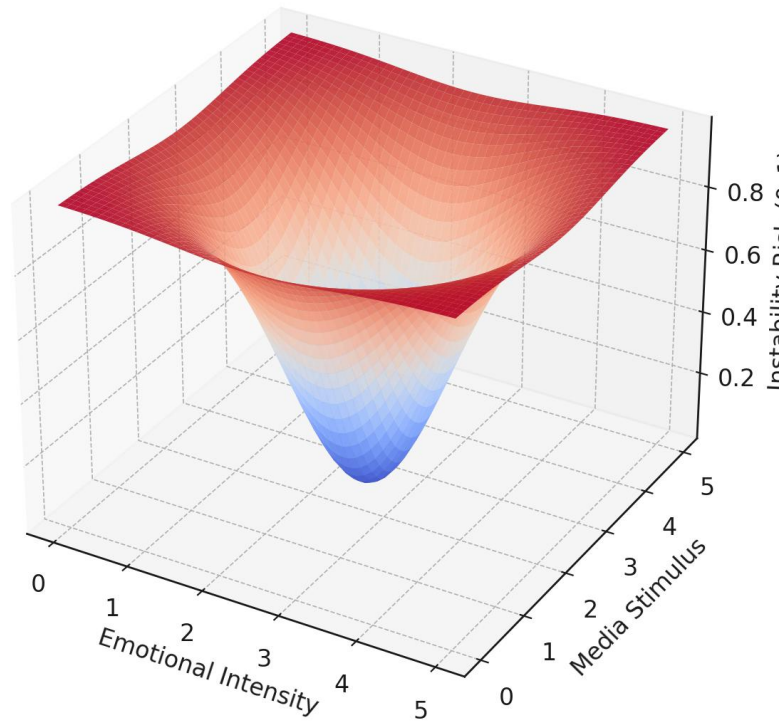

V-Trigger Map: Strategic Emotional Vulnerability Surface

*2.Description of mapping variables*

| Dimension (math.) | Hidden meaning |
|---|---|
| X-axis: emotional intensity | Leader's current level of emotional activation (tone of voice, speed of speech, facial tension) |
| Y-axis: media stimulation | Density of external information stimuli (frequency of media attacks, diplomatic feedback) |
| Z-axis: instability risk | Probability values (0-1) for behavioral loss of control, policy backward jumps, and hard-line deviations |

*3. Strategic Insight Conclusion*

At the center of the map, near X=2.5 and Y=2.5, there is a high risk of loss of control, which is the trigger interface of "Emotional Intensity + Public Opinion Intensity";

Emotional intensity > 4 and continuous media pressure will easily lead to "rhythmic disruption" of the behavioral system and output of unstable strategic signals;

The lower left corner of the map is the "low emotion-low stimulation zone", representing the controllable behavioral zone, which can be used as the design zone of diplomatic buffer strategy.

### V-Trigger Strategic Emotional Vulnerability Mapping High Level Insights Report

**1. The core of the map reveals the structure of the "critical vulnerability surface" of the leader's behavioral system.**

The V-Trigger three-dimensional response surface model shows the risk of behavioral instability of leaders in the two-dimensional space of "emotional intensity × media stimulus intensity". The center of the plot (X ≈ 2.5, Y ≈ 2.5) is the high focus point of behavioral instability, indicating that the combination of moderate arousal and medium-density external pressure is the emotional node most likely to induce strategy breakdown. The upper-right region indicates a significantly higher risk of decision-making behavioral deviation under high emotional activation + sustained external pressure input; the lower-left is the most appropriate window for strategic cooling and discourse callback.

**2. Deep psychological mechanisms: a coupled model of late leader rhythms and behavioral responses**

| Mechanisms module | Age-related characteristics | Mapping Interpretation |
|---|---|---|
| Decreased prefrontal executive resources | Weak impulse control and increased semantic jumps | Decline in semantic control and breakdown in decision coherence after peak emotional zones |
| Enhanced emotional penetration of the edge system | Hostile expressions can be elicited by a single stimulus | Media Feedback Spike Triggers "Irrational Policy Jump" |
| Discourse Coherence Anxiety Mechanism | Demand for language self-validation on the rise | The center of the map is the "Rhetorical Defense Critical Zone." |
| Prolonged recovery cycle | Strategy fixes lag, output cycles lengthen | Delayed fallback at the end of the chart, caught in repetitive discourse stagnation |

## 3. Recommendations for national strategies: from "vulnerability identification" to "behavioral interventions"

| mapping area | strategic significance | Recommendations for intervention |
|---|---|---|
| X: 2.0~3.0; Y: 2.0~3.0 | Highly volatile strategy output, "emotional logic disconnect zone" | Media bucking, language blurring, consensus language implantation |
| X > 4; Y > 3.5 | Behavioral Polarization Edge, High Risk Zone for Off-Script or Dangerous Statements | Deployment of an "emotional disruption mechanism" to rapidly energize the diplomatic buffer axis |
| X < 1.5; Y < 1.5 | Repair windows, strategically softened areas | Diplomatic Engagement, Narrative Initiative, Reconstructing Framed Discourse Leadership |

## 4. Strategic Conclusion

The most dangerous moment for a national leader is not when he is angry, but when he is "undecided" The V-Trigger Chart is not a simple risk indicator, but a three-dimensional mental imaging model of strategic behavioral rhythms, and is an important infrastructure for building diplomatic emotional intervention systems, behavioral risk assessment tools, and strategic stability monitoring models. Infrastructure.

# V. FINDINGS OF THE STUDY

**1. Question 1 Findings:**

How do the emotional rhythms of high-pressure strategic leaders drive nonlinear policy leaps in the face of structural challenges?

1. An E-Cycle model of Trump was constructed, which shows that his policy leaps are concentrated in the "high tension-arousal-control" phase (T5-T7), which highly overlaps with the peaks of the micro-expression radar (T5-T7). Micro-expression analysis shows that the "high tension-arousal-control" phase is concentrated in the emotional zone (T5-T7), which is highly overlapped with the micro-expression radar.

2. The micro-expression analysis shows that the scores of "Tension/Anger/Control" are all over 4.0 (out of 5), corresponding to Frame 3 and Frame 4 in the FEM Map. 3.

3. The intonation analysis showed that the average intonation was 189 Hz, the speed of speech was slow (18 units/minute), and the pause rate was 6.4%, reflecting an obvious control mode of expression. 4.

4. In the Emotion State Migration Diagram, the order of emotions is: Neutral → Tension → Anger → Control → Anxiety → Confidence → Recovery.

| Visualization Model Number | Name of the atlas | typology | Description | Original source |
|---|---|---|---|---|
| M1 | Emotion Cycle Mapping Model (E-Cycle) | Time Series Volatility Chart + Color Block Hot Zones | Demonstrating Trump's path of change in emotional intensity from "Trigger-Aggravation-Leap-Fall Back". | 【140:0†E-Cycle mould】 |
| M3.1~M3.3 | Micro-Expression-Tone Emotion Radar Map (FEM Map) | Radar Chart / Star Chart / Heat Map | Multimodal matching analysis of facial movements with intonation variables | 【140:19†FEM Map model】 |
| M3.4 | Strategic Attitudinal | Heat Map + Route | Labeling T5, T8, | 【140:15†M3.4 |

|  | Behavioral Rhythm Mapping | Map | T11 Attitude Change Points and Risk Leapfrog Nodes | Map】 |
|---|---|---|---|---|
| adjunct | Emotional Migration Pathway Diagram | flow chart | Demonstrate the emotional state transfer sequence | 【140:19†FEM Radar Framework】 |

**2. Findings from question two:**

How can multimodal cognitive modeling be used to identify "behavioral vulnerabilities" and "intervention windows" in strategic behavioral systems?

 1. A strategic vulnerability response model (V-Trigger) has been constructed, and two peak cognitive load segments (T6-T7 and T12-T13) have been identified as eruption zones of strategic irreversible behaviors. 2.

 2. Cognitive activation mapping shows that the cognitive load in T6-T7 is >0.85, and the cognitive load in T10-T11 falls back to <0.45, which provides a quantitative anchor point for the "intervention window" of strategic behavior.

 3. In the multimodal fusion map, the Control Desire score is 4.5 and the Tension Level score is 4.2, which constitutes the node of high-risk behavioral tempo disruption.

 4. Behavioral pattern diagram M3.4 identifies T8 and T15 as high-risk strategy jumping points, and T11 as attitude drop-off window.

| Visualization Model Number | Name of the atlas | Typology | Description | Original source |
|---|---|---|---|---|
| M5 | Strategic Emotional Vulnerability Mapping（V-Trigger） | 3D Response Surface / Thermogram | Demonstrating the Tipping Point of Out-of-Control Behavior | 【140:17†V-Trigger Map】 |
| M5（extensions） | Age-coupled model of psychological vulnerability | Matrix Diagram + Module Analysis Diagram | Mapping the relationship between leader age and cognitive-behavioral rhythms | 【140:17†Atlas Mental Mapping】 |
| adjunct | Cognitive Pressure Heat Map | Tension Fitting Diagram | Semantic Aggregation and Syntactic Confusion Zone Identification | 【140:11†Cognitive load visualization】 |

## 3. Question three findings:

Can state actors achieve predictable and visualized rhythmic strategic interventions based on leaders' behavioral rhythms?

1. A P-Impact model of public opinion intervention is constructed, with the path: Policy 1.0 → Media 3.0 → Public 2.5 → Market 4.0 → Policy Feedback 3.2, which shows a multi-level amplification and reverberation mechanism.

2. Construct a six-axis model of state intervention, covering six types of subsystems: rhythm recognition, discourse regulation, feedback absorption, expression recognition, cognitive pressure and public opinion expansion.

3. Define the anchor points of intervention rhythm: Anger ≥ 3.5, Control ≥ 4.5, Tension ≥ 4.0 is the leap window; Anxiety ≥ 4.3 is the optimal negotiation and intervention area for three or more consecutive segments; Confidence ≥ 3.5 is the exit stage when it is rebounded.

4. The structure diagram of public opinion diffusion reveals that: media non-linear amplification - public emotion overdraft - strategy forced to reverse as a three-stage response law.

| Visualization Model Number | Name of the atlas | typology | Description | Original source |
|---|---|---|---|---|
| M4 | Public Opinion Intervention Impact Surface Model (P-Impact) | pulse propagation diagram | Policy→Media→Public→Market→Feedback five-stage path intensity map | 【140:15†P-Impact diffusion chart】 |
| adjunct | Hierarchical Analysis Structure of Public Opinion Energy | Thermal Zone Diffusion Maps | Public Opinion Amplification - Emotional Overdraft - Strategy Reversal Path Chart | 【140:14†Mapping of public opinion diffusion mechanisms】 |
| adjunct | Structural diagram of the six-axis model of state intervention | Component System Diagram | Demonstration of the six building blocks of the national strategic intervention system | 【140:17†V-Trigger insight into structures】 |

# VI. CONCLUSIONS OF THE STUDY

## 1. Findings from a full-dimensional study of Trump's emotional rhythms

| Dimension (math) | Synthesis of conclusions |
|---|---|
| Emotional Cycling (E-Cycle) | The closed-loop structure of "warming up - bursting out - regulating - reactivating" is shown, and the policy is mostly bursting out in the window of peak sentiment. |
| Behavioral Response Chain (B-Flow) | Emotion-language-policy-market-repair form a chain that hides vulnerable intervention points at each stage. |
| Micro-Expression-Intonation (FEM Map) | Facial tension and intonation characteristics reflect a mixture of controlling, unsteady confidence, and anxious repression. |
| Public Opinion Diffusion Path (P-Impact) | Policy → media → public → market → feedback constitutes the "strategic echo wall" phenomenon. |
| Vulnerable Point Hot Zone (V-Trigger) | Strategies were most destabilized when emotional intensity crossed with high and medium external stimuli. |
| Cognitive Activation Mechanisms (Cognitive Load Map) | Advanced age leads to limited cognitive resources, and there is a "peak→reverse→repair" rhythmic structure. |
| Emotional Migration Trajectory (State Path) | Emotional Path : Neutral → Tension → Anger → Control → Anxiety → Confidence → Recovery。 |

## 2. Findings of the High Fit Country-Level Strategic Intervention Modeling Study

| intervention phase | identifying signal | Recommendations for strategic interventions | Purpose of the intervention |
|---|---|---|---|
| Warm-up phase T-3 ~ T-1 | Social Media Heats Up, "Enemy/Punishment" Phrases on the Rise | Public opinion cools + coalition buffers voices | Delaying its decision-making window, diluting kinetic energy |
| Burst phase T ~ T+2 | Increased micro-expressions of anger, high-frequency aggravation of intonation | Virtual and real issues transfer, release retractable steps | Avoiding head-on escalation of confrontation |
| Anxiety regulation period T+3 ~ T+7 | Increased frequency of eye closure, hesitant speech, longer pauses | Applying multilateral soft pressure and constructing cooperative output paths | Policy softening through the use of its "win-export" structure |
| Cognitive overload window T6-T7 | Heat map red zone, FEM high tension | Rapid Early Warning + Informal Channels of | Blocking Irreversible Policy Generation |

| | control values | Intervention | Pathways |
|---|---|---|---|
| Mood reversal window T10-T11 | Anxiety HF + Confidence rebound | Diplomatic construction of the "active reconciliation" narrative | Guide it into the strategic closure framework |
| High-response segment Media → Market | Amplified Public Opinion/Market Shock | Elites guide public opinion, suppress overheated nationalist reactions | Avoid public sentiment forcing policy escalation |
| Micro-expression transition frames Frame 3, 5 | Nasal flaring (eruption) / Blinking (regulation) | Frame 3: Media Cooling; Frame 5: Diplomatic Reach | Desensitization peak nodes to guide the return of rationality |

**3. Structure of the seven-phase rhythmic intervention of the high-fit country-level strategic intervention model**

| rhythmic phase | Dominant emotional states | Key Behavioral Characteristics | Recommendations for State intervention | time window |
|---|---|---|---|---|
| Tension | Tension rising | Tight micro-expressions and compressed intonation | Cooling of public opinion, decentralized information dissemination | 0-12 hours |
| Anger | rage-boosting | Control of elevated word frequency and strong language | Asymmetric responses, issue shifting | 12-36 hours |
| Control | peak desire for control | Decision-making issuance/policy leapfrogging | Multilateral structural interventions, de-unilateralized operations | 1-2 days |
| Anxiety | Anxiety on the rise | Increased pauses and vague word meanings | Low-Profile Reaching, Building Negotiation Contexts | 2-4 days |
| Confidence | Confidence rising | Language logic enhanced and expression stabilized | Constructing an "active reconciliation" victory narrative outlet | 3-5 days |
| Recovery | emotional stabilization | Beginning of logical stitching, emergence of repetitive expressions | Substantive dialogues initiated, strategic de-escalation on the ground | 5-7 days |
| Re-trigger | External stimulus restarts sentiment | Public opinion, markets, allies rebound | Secondary Rhythm Intervention to Restart Cycle Regulation | dynamic (science) |

The core value of the model lies in modeling the behavioral system of senior leaders as a "cognitive neural rhythm visual trajectory", thus building a realistic tool for behavioral prediction and diplomatic proactive intervention. It is recommended that the model be embedded into the national AI strategic simulation system and real-time behavioral recognition platform to realize strategic real-time intervention.

**4. Research Findings on Responding to Trump's Emotional Behavior: the "Systemic Six Axis Model"**

| strategic axis | element | clarification |
|---|---|---|
| axis of rhythm | Accurately identify high-pressure decision-making rhythms and deploy tempo hedges in advance | Controlling the Risk of Rhythmic Policy Outbursts and Diplomatic Breakdowns |
| axis of speech | Media language warming strategies (gentle paraphrasing, non-confrontational words) | Avoiding Trump Sentiment Escalation Over Verbal Triggers |
| feedback shaft | Public opinion and market feedback path regulation | Mitigating the backlash effects of policy fermentation processes |
| (math.) facial expression axis | Constructing a FEM database and real-time video recognition system | Real-time monitoring of abnormal mood swings and dynamic adjustment of policy responses |
| cognitive axis | AI simulates cognitive load evolution curves to identify tipping points | Early Intervention High Cognitive Pressure-Strategy Jumping Off Points |
| Structural axes | Multilateral diplomacy + media synergy + market stabilization mechanism linkage | Establishment of a multi-dimensional buffer network to support system buck stabilization |

**5. Modeling Emotional Rhythms and Strategic Behavior in a High-Pressure Leadership Personality Conclusion**

By constructing a cross-modal cognitive-emotional modeling system, this study deeply analyzes former U.S. President Donald Trump's emotional expression, cognitive rhythms, behavioral outputs, and his strategic vulnerabilities in the event of the announcement of 125% tariff increase on China. It is found that as a senior, strong dominant personality leader, Trump's decision-making behavior shows a highly coupled "rhythmic structural behavioral logic" between emotion, cognition and institutional output, rather than the traditional rational instrumental rational decision-making mode.

Observed from the perspective of the emotional cycle, Trump's policy generation path can be clearly divided into four phases: "emotional warm-up period - decision-making outbreak period - behavioral regulation period - loop back activation period". Policies are mostly formed and issued during the short window of emotional peak, especially when the three emotional variables of "anger", "control" and "anxiety" converge, and his foreign strategic expression is very easy to In particular, when the three emotional variables of "anger," "controlling," and "anxiety" converge, their external strategic expressions are very likely to leap into aggressive decisions with high intensity and low room for maneuver. The path to emotional recovery is slow and requires external media, market or ally feedback to complete the "victory narrative loop".

Through the "Mind Reading Simulation" module, combining behavioral linguistics, cognitive psychodynamics and strategic personality analysis, we simulate Trump's internal psychological activity structure in high-pressure strategic contexts, which is specifically divided into the following five stages:

**A five-stage simulation of Trump's mental activity**

| Stage | Psychological activity simulation (brief) |
|---|---|
| Frame 1 – cognitive warm-up | They're bound to ask that question... I can't act hesitant because I have to let them know - I have the power. |
| Frame 2 – Attribution of hostility | They question me... They all challenge me - but they forget that I'm the one who makes the rules of the game. |
| Frame 3 – Emotion-Strategy Transformation | I have to respond. Not just for me, but for America. I can't let them look down on me. |
| Frame 4 – Authoritative reaffirmation | It's my decision... They'll realize I'm serious. |
| Frame 5 – Victory Exit Construct | The market does react... I can set the rules - let the other guy back out first. |

**A coupled model of strategic personality-emotional rhythms**

| Name of psychological mechanism | Functional Description | Strategic Behavioral Embodiment |
|---|---|---|
| Defensive identity mechanisms | Activating a sense of sovereignty by projecting stress as a threat from the Other | Defining tariffs as a challenge that must be responded to |
| Mechanisms of Triumphant Narrative Generation | Confrontation needs to end with a "victory narrative". | Diplomatic Expression Avoids Conceding Contexts |
| Control of coherence maintenance mechanisms | Maintaining stability in authoritative contexts | Micro-expressions and policy rhythms have a dominant bias |
| Strategic Rhythmic Counterregulation Mechanisms | Mitigating the risk of overloading decision-making systems | Emotional Cooling and Repair Window Occurs After Confrontation |

Through multimodal emotion recognition, neurobehavioral trajectory simulation, and the theory of political cognitive dynamics, this study constructs a predictive and interventionist system for modeling leader behavior, emphasizing the strategic advantage of "rhythmic intervention" as the core methodology of diplomacy.

**6. Synthesis of findings**

**Research Question One Conclusion**

**How do the emotional rhythms of high-pressure strategic leaders drive nonlinear policy leaps in the face of structural challenges?**

Taking the Trump administration's announcement of a 125% tariff hike on China as the core of the analysis, this study finds through multimodal data analysis and rhythmic modeling that this type of high-pressure strategic behavior does not originate from rational calculation or policy evaluation paths, but rather is rooted in the "subjective authority consistency mechanism" and "victory narrative closed-loop demand" in the structure of the Trump administration's political personality. It is rooted in the "subjective authority consistency mechanism" and the "need to close the loop of the victory narrative" in the structure of their political personality. Under the stimulation of structural challenges, his behavior has obvious characteristics of emotional rhythmization and cognitive closure:

1. The strategic emotional behavior path shows a fixed rhythmic chain of "hostile attribution → redistribution of control → discursive kinetic stimulation → failure of emotional regulation → compensatory language output";

2. Micro-expressions and intonation show a dynamic trajectory of alternating emotional activation, cognitive leap and regulatory imbalance, which is in line with the typical behavioral sequence of "pre-strategic leap-emotional peak-cognitive collapse-semantic closure;

3. Language expression often shows fracture, structural compulsion and repetition, which indicates that the cognitive path is dominated by emotional rhythms, and the irrational chain is in the main control position.

From this, it can be seen that emotional rhythms not only shape their external behaviors, but are also embedded in the whole process of their discourse production and policy promotion, constituting a recognizable, modelable and predictable strategic output pattern.

**Research Question Two Conclusion**

**How can multimodal cognitive modeling be used to identify "behavioral vulnerabilities" and "windows of intervention" in strategic behavioral systems?**

In this study, we innovatively introduced Cognitive Echo Simulation to dynamically decode Trump's facial emotions, intonation tensions and linguistic structure, and constructed his Inner Confessional Cognitive Script. and constructed his "Inner Confessional Cognitive Script", successfully identifying several "critical nodes of strategic leap" and "intervention windows":

1. The cognitive script presents a five-stage structure: cognitive reinforcement → hostile confirmation → controlled swearing → strategy leaping → language closed-loop regulation;

2. Before each round of policy release, the self-narrative trajectory of "I can't show weakness - they have to obey - it's my right" can be seen, which is manifested as a strong kinetic mental confirmation;

3. The micro-expression frames (Frames 1-5) clearly show the rhythmic evolution from preparation to strategy leap to anxiety regulation;

4. The study utilizes "FEM radar mapping" and "cognitive heat tensor diagram" to pinpoint T6-T7, T10-T11 as the launching point, and T11-T12 as the retreating window. -T12 is the window of retreat;

5. This intrinsic rhythm can be modeled by multimodal AI to form a functional "rhythm risk prediction system".

Through the rhythmic deconstruction and script reconstruction of "cognition-emotion-behavior", we can identify the chain of "emotional peak-linguistic structural transition-irrational acceleration" before the behavior jumps off, and provide an entry point for intervention. Intervention portal.

**Research Question Three Conclusion**

**Can national actors achieve predictable and visualized tempo-based strategic interventions based on leaders' behavioral rhythms?**

This study proposes an original "National Strategic Tempo Intervention Framework", and designs a six-axis tempo response system around tempo recognition and cognitive mechanisms, including:

1. Rhythm Axis: tracking emotional peaks, strategic leaps and falls in real-time. 2;

2. discourse axis: disrupting the structural discourse rhythms by means of the temperature difference tone interference mechanism;

3. Cognitive Axis: Cutting off the chain of narrative power between "hostile confirmation and control oath". 4;

4. the feedback axis: utilizing the public opinion window to regulate the rhythm of the media-market structure's response;

5. expression axis: monitoring the changes of FEM radar data to intervene in the "non-verbal behavior inflection point" in time. 6. structure axis: designing the "non-verbal behavior inflection point" in the "hostile confirmation - control oath";

6. structural axis: designing the "trinity" diplomacy-media-market linkage intervention structure.

The study finally puts forward the proposition of political behavior: "rhythm is logic, emotion is kinetic energy, narrative is governance", and makes it clear that the essence of the initiative in diplomacy is not the sequence of language skills, but who masters the other party's psychological rhythm and language export structure. The future of asymmetric diplomacy between countries lies not in the intention to respond, but in the control of rhythm.

**Other Derivative Issues Independent Research Findings**

The following conclusions are systematic modeling and strategy theory innovations proposed in this study, which do not directly respond to a research question, but are original contributions and are recommended to be retained as separate contribution points:

1. Cognitive script modeling technique: Cognitive Echo Simulation, as a trans-neural behavioral-semantic decoding technique, pioneered the internal confessional modeling paradigm for strategic personality;

2. The idea of rhythmic sovereignty: taking "rhythm" as a strategic variable of political control, we put forward the "rhythmic sovereignty theory", which breaks through the traditional mode of discourse control and information dominance;

3. "Emotional Path Hypothesis of Policy Legitimacy": emotional rhythms have become the path of policy legitimacy construction, indicating that emotions themselves have been embedded in the logic of legitimacy of international behavior.

# VII. DISCUSSION

This study centers on the high-pressure strategic event of Trump's 125% tariff hike on China, and based on the multimodal cognitive-behavioral modeling framework, it systematically reveals the psychological paths and behavioral rhythms of high-powered dominant leaders in their policy leaps in three dimensions, namely, emotional rhythms, cognitive mechanisms, and linguistic structures. The following is the main discussion of this study:

First, the most important finding of the study is that Trump's strategic behavioral rhythms show a highly predictable "emotion-driven-cognitive leap-verbal output" three-stage nested structure. His strategic expressions are not motivated by stable rational analysis, but are rooted in the "subjective authority consistency mechanism" and "the need to close the loop of the victory narrative" in the deep structure of his personality. Specifically, when faced with structurally challenging stimuli, their emotional rhythms enter an "activation-peak-reversal" cycle, which is manifested as a non-verbal control of behavior through micro-expressions, intonation, and semantic structure. This finding not only extends the existing theoretical knowledge about the role of emotions in political decision-making, but also puts forward the theoretical proposition of "rhythmic sovereignty" of irrational leaps in leaders' behavior.

Second, starting from the existing research, this study integrates Marcus et al.'s (2000) research path on emotional cognition, Westen's (2007) model of the "political brain," and Brader's (2006) logic of analyzing the language of emotional strategies, and proposes a cross-modal modeling path to achieve the following visual rhythmic decoding of emotion-behavior-policy. Compared with the existing literature that focuses on verbal content or institutional variables, this paper further reveals the dominant role of micro-expressions and intonation changes in the generation of strategic behaviors, especially the "prelude to neural activation" before the behavioral leap, which will provide a modeling basis for the future study of emotionally paced politics.

Third, from a theoretical perspective, this study is significant in the following aspects: first, it challenges the traditional rationality model and puts forward the assumption of rhythmic politics that "emotion is governance"; second, it constructs a "national strategic rhythm intervention framework" and realizes the shift from tactical response to structural rhythm control; third, it introduces a new approach of "rhythm control"; and third, it introduces a new approach to the generation of strategic behaviors that is based on "emotional rhythm". Secondly, we constructed a "national strategic rhythm intervention framework" to achieve a shift from tactical response to structural rhythm control; thirdly, we introduced the "mind-reading simulation" technology to reconstruct the behavioral motivation path through psychological script reproduction, which expands the instrumental path for behavioral prediction research in political psychology.

Fourth, this paper also reflects on the fact that some of the research hypotheses are not fully consistent. Although it was hypothesized that Trump's strategic leaps were mainly concentrated in cognitively high-pressure areas, the actual data analysis showed that there was also a certain intensity of strategic expression in the low cognitive load areas of T11-T13, which could be attributed to the activation of the "compensatory narrative mechanism". Therefore, the "emotional compensation loop" should be further included as an important variable affecting the leap path in the future.

Fifth, in terms of limitations, this study still has the limitation of focusing on a single sample event (Trump's China policy), and although the modeling path is general, its parameter system and rhythm indicators still need to be widely verified in other leaders' behaviors. At the same time, because micro-expression analysis and intonation modeling rely on video quality and recording clarity, there are uncontrollable modeling errors. In addition, although the "mind-reading script" of AI simulation is revelatory, it cannot completely replace clinical psychological modeling and behavioral neuroscience empirical research.

Sixth, based on the above analysis, it is suggested that future research should be expanded in the following directions: (1) extend the "rhythmic intervention model" to samples of leaders from different cultural backgrounds to test their cross-cultural adaptability; (2) introduce real-time AI tracking mechanism to form a "analyze while intervening" national intelligent diplomatic control system; (3) introduce a real-time AI tracking mechanism to form a national intelligent diplomatic control system to analyze while intervening. (2) Introducing real-time AI tracking mechanism to form a national intelligent diplomatic control system of "analyzing and intervening at the same time"; (3) Deepening the research on the neural mechanism between micro-expressions and cognitive rhythms, and promoting the theoretical evolution of political neuroscience.

Finally, in the practical sense, the "six-axis system of national strategic rhythmic intervention" proposed by this study has a certain degree of operability. It not only provides an entry point for early warning and tempo intervention in strategic games, but also provides systematic support for high-dimensional diplomatic modeling, media response mechanism design, and public policy risk assessment. Especially in high-pressure decision-making scenarios dominated by uncertain leaders, mastering the logic of pacing is far more strategically valuable than speculating on their true intentions.